\documentclass[preprint,showpacs,aps,amsmath,amssymb,
nofootinbib,showkeys]{revtex4}

\usepackage{epsfig}
\usepackage{subfigure}
\usepackage{amsmath}
\usepackage{wasysym}
\usepackage{bbm}

\def\beq{\begin{equation}}
\def\eeq{\end{equation}}
\def\barr{\begin{eqnarray}}
\def\earr{\end{eqnarray}}
\newcommand{\beqa}{\begin{eqnarray}}
\newcommand{\eeqa}{\end{eqnarray}}

\def\gsim{\:\raisebox{-1.1ex}{$\stackrel{\textstyle>}{\sim}$}\:}

\def\dma{\Delta m^2_{\rm atm}}
\def\dms{\Delta m^2_{\odot}}
\def\dmsq{\Delta m^2}
\def\dmst{\Delta m^2_{\rm st}}
\def\dd{\Delta_{32}} 
\def\nn{\nonumber}

\def\mubar{\bar{\mu}}
\def\taubar{\bar{\tau}}
\def\nn{\nonumber}
\def\nbar{\overline{N}}

\begin{document}

%\rightline{\today}

\title{Signatures of heavy sterile neutrinos at 
long baseline experiments}

\author{Amol Dighe}
\email{amol@theory.tifr.res.in}
\affiliation{Tata Institute of Fundamental Research, \\
Homi Bhabha Road, Colaba, Mumbai 400005, India}

\author{Shamayita Ray}
\email{shamayitar@theory.tifr.res.in}
\affiliation{Tata Institute of Fundamental Research, \\
Homi Bhabha Road, Colaba, Mumbai 400005, India}

%%%%%%%%%%%%%%%%%%%%%%%%%%%%%%%%%%%%%%%%%%%%%%%%%%%%%%%%%
\begin{abstract}

Sterile neutrinos with masses $\sim 0.1$ eV or higher
 would play an important role in astrophysics and
cosmology.
We explore possible signatures of such sterile neutrinos
at long baseline experiments.
We determine the neutrino conversion probabilities
analytically in a 4-neutrino framework, including
matter effects, treating
the sterile mixing angles $\theta_{14}, \theta_{24}, 
\theta_{34}$, the deviation of $\theta_{23}$ from maximality,
as well as $\theta_{13}$ and the ratio 
$\dmsq_\odot/\dmsq_{atm}$
as small parameters for a perturbative expansion.
This gives rise to analytically tractable expressions
for flavor conversion probabilities from which effects
of these parameters can be clearly understood.
We numerically calculate the signals at a neutrino factory 
with near and far detectors that can identify the 
lepton charge, and point out observables that can 
discern the sterile mixing signals.
We find that clean identification of sterile mixing
would be possible for  $\theta_{24}\theta_{34} \gsim 0.005$ 
and $\theta_{14} \gsim 0.06$ rad with the current 
bound of $\theta_{13} < 0.2$ rad; a better $\theta_{13}$ 
bound would allow probing smaller values of sterile mixing.
We also generalize the formalism for any number of 
sterile neutrinos, and demonstrate that only certain
combinations of sterile mixing parameters are relevant
irrespective of the number of sterile neutrinos. 
This also leads to a stringent test of the scenario
with multiple sterile neutrinos that currently is able
to describe all the data from the short baseline
experiments, including LSND and MiniBOONE.
\end{abstract}



\preprint{TIFR/TH/07-21}

\maketitle

%%%%%%%%%%%%%%%%%%%%%%%%%%%%%%%%%%%%%%%%%%%%%%%%%%%%%%%%%%%%%
\section{Introduction}\label{intro}
%%%%%%%%%%%%%%%%%%%%%%%%%%%%%%%%%%%%%%%%%%%%%%%%%%%%%%%%%%%%

In the framework of the Standard model (SM) 
of particle physics, 
there are only three neutrinos, one each with the
electron, muon and tau flavor. 
The LEP experiments have determined the 
number of light neutrinos that couple with the 
Z boson through electroweak interactions to be 
$2.984 \pm 0.008$ \cite{lep}, 
thus closing the door on any more generations
of ``active'' neutrinos. 
However, there still may exist sterile neutrinos that
do not have electroweak interactions. 
Though they cannot be detected in the $Z$ decay,
they may mix with the active neutrinos and hence
participate in neutrino oscillations. 
The extent of sterile neutrino participation in
solar neutrino data is severely restricted from the
neutral current data from SNO \cite{sno-nc}.
The atmospheric neutrino data show that the major
contribution to the muon neutrino disappearance
has to be from $\nu_\mu \leftrightarrow \nu_\tau$ 
oscillations, however a small admixture of sterile
neutrinos cannot be ruled out \cite{atm-sterile,atm-st-quant}.
Short baseline experiments sensitive to sterile
neutrinos in the $\sim 1$ eV range 
\cite{dydak,bugey,karmen,nomad} have given strong
upper bounds on sterile mixing. 
The CNGS experiment expects to further restrict the 
sterile mixing parameter space through the
$\nu_\mu \to \nu_{e,\tau}$ channels \cite{cngs}.
The MiniBOONE experiment \cite{miniboone} has
virtually ruled out any effect of sterile neutrinos
in the LSND parameter space \cite{lsnd} 
if there were only one sterile neutrino species. 
However, recently it has been pointed out 
\cite{maltoni-schwetz}
that with two or more sterile neutrinos, it is 
possible to be in agreement with all the data.

Even if the LSND results are ignored, so that
there is no longer any need for sterile neutrinos
for explaining the neutrino oscillation data,
sterile neutrinos that obey all the constraints
from the terrestrial experiments can still play a crucial
role in astrophysics and cosmology \cite{kusenko}.
The matter enhanced active-sterile neutrino transformation 
can have a great effect on r-process nucleosynthesis 
in the core-collapse supernovae
\cite{r-process},
and can also influence the explosion dynamics 
\cite{sn-explosion}.
Moreover, the anisotropy inside the exploding supernova
may be transported outside efficiently by sterile neutrinos,
thus helping to explain the large observed velocities 
of pulsars \cite{pulsar-kicks}.
Sterile neutrinos of mass $\sim$ keV are also
excellent dark matter candidates: $\nu$MSM (SM with
three sterile neutrinos) \cite{nu-msm} manages to explain 
masses of active neutrinos, baryon asymmetry of the universe 
and the abundance of dark matter together. 
The Chandra blank sky observations also allow 
keV neutrinos to be viable dark matter candidates
\cite{chandra}.
Such heavy dark matter also helps in the production of 
supermassive black holes \cite{supermassive}.
Sterile neutrinos may leave their imprints in the
supernova neutrino burst \cite{sn-choubey1,sn-choubey2}, 
or in the ultrahigh energy neutrino signals 
observed at the neutrino telescopes \cite{uhe-choubey}.

The main requirements for the astrophysically and
cosmologically relevant sterile neutrinos are thus
that they be heavy ($m \sim$ 1--10 eV  
for r-process nucleosynthesis, and $m \sim$ keV 
for the dark matter candidates) 
and that they mix weakly with the electron 
and muon neutrino
(in order to satisfy the MiniBOONE constraints).
In this article, we consider the case with one
such sterile neutrino, which may influence 
neutrino oscillation experiments.
We perform the complete 4-$\nu$ analysis,
taking into account all the three additional
mixing angles of the sterile neutrino $\nu_s$ 
with the active ones, and the two additional
CP violating phases.
We only concentrate on heavy neutrinos, such that
if $m_4$ is the mass of the neutrino eigenstate 
with a dominant sterile component, 
$\dmst \equiv |m_4^2 - m_i^2| \gsim 0.1$ eV$^2$ 
for all other neutrino mass eigenstates $\nu_i$.
The oscillations due to $\dmst$ are
rather rapid, and can be taken to be averaged out
in the long baseline data.
As a result, the data are expected to be
insensitive to the exact value of $\dmst$.
However, the additional mixing angles $\theta_{i4}$
may leave their signatures in the data.

We treat the effects of the sterile neutrino as a 
perturbation parametrized by a small auxiliary 
parameter $\lambda \equiv 0.2$. 
To this end, we represent 
the active-sterile mixing angles 
$\theta_{14},\theta_{24},\theta_{34}$, 
the deviation of $\theta_{23}$ from
maximality, the reactor angle $\theta_{13}$
as well as the ratio $\dms/\dma$, formally as
some power of $\lambda$ times ${\cal O}(1)$ numbers, 
so that a systematic expansion in powers of $\lambda$ 
may be carried out.
Averaging out the fast oscillations due to $\dmst$
allows us to obtain simple analytic approximations for 
the flavor conversion probabilities of neutrinos.
The expressions thus obtained describe the dependence 
of relevant conversion or survival probabilities on 
the parameters in a transparent manner.

We analyze, using analytical as well as numerical means, 
how the parameters involving sterile neutrinos --
constrained by the data from solar, atmospheric, and short
baseline experiments --
affect the results at the long baseline experiments.
We illustrate this effect quantitatively in the case of
a neutrino factory setup involving a near and a far
detector that are capable of lepton charge identification.
In particular, we consider the CP asymmetry in $\mu$ and 
$\tau$ channels as the observables and calculate 
how far the limits on the sterile mixing parameters can be 
brought down. 
We also consider the electron channel, where 
signals of sterile
neutrino mixing can still be established by the 
counting of the total number of events above a threshold.

The paper is organized as follows. 
In Sec.~\ref{expansion}, we explain our formalism
of a systematic expansion of all quantities in 
an auxiliary small parameter $\lambda$ and the
use of perturbation theory to obtain the neutrino
flavor conversion probabilities.
In Sec.~\ref{lbl} we examine some of the possible 
signatures of sterile neutrino mixing on the signals
at a neutrino factory setup with near and far detectors,
where we also estimate bounds that can be obtained
at such long baseline experiments.
In Sec.~\ref{many-sterile}, we generalize our formalism to
any number of sterile species and point out that only
certain combinations of the sterile mixing parameters
are relevant, independent of the number of sterile species.
Sec.~\ref{concl} concludes.

%%%%%%%%%%%%%%%%%%%%%%%%%%%%%%%%%%%%%%%%%%%%%%%%%%%%%%%%%%%%%
\section{Analytic computation of neutrino flavor 
conversion probabilities}
\label{expansion}
%%%%%%%%%%%%%%%%%%%%%%%%%%%%%%%%%%%%%%%%%%%%%%%%%%%%%%%%%%%%%

We work in the 4-$\nu$ framework, where $(\nu_e,\nu_\mu,
\nu_\tau,\nu_s)$ form the basis of neutrino flavor eigenstates
and $(\nu_1,\nu_2,\nu_3,\nu_4)$ form the basis of
neutrino  mass eigenstates. 
The mass eigenstates are numbered according to the 
convention 
$|\dmsq_{42}| \gg |\Delta m^2_{32}| \gg \Delta m^2_{21} >0$,
where $\dmsq_{ij} \equiv m_i^2 - m_j^2$.
We have $\dmst \approx |\dmsq_{42}|, \dma \approx |\dmsq_{32}|$
and $\dms \approx \dmsq_{21}$.
Note that the sign of $\dmsq_{32}$ is as yet unknown,
a positive (negative) $\dmsq_{32}$ corresponds to
the normal (inverted) mass ordering of neutrinos.

The mass and flavor eigenstates of neutrinos
are connected through a unitary matrix ${\cal U}$, such that 
\beq
\nu_\alpha = {\cal U}_{\alpha i} \nu_i \; ,
\eeq
where $\alpha \in \{e, \mu, \tau, s \}$ 
and $i \in \{1,2,3,4\}$.
The mixing matrix ${\cal U}$ may be parametrized as
\beq
{\cal U} = 
U_{14}(\theta_{14}, \delta_{14}) \;
U_{34}(\theta_{34},0) \; 
U_{24}(\theta_{24}, \delta_{24}) \; 
U_{23}(\theta_{23},0) \;
U_{13}(\theta_{13}, \delta_{13}) \;
U_{12}(\theta_{12}, 0) \; ,
\label{u-general}
\eeq
where  $U_{ij}(\theta_{ij},\delta_{ij})$ is the complex rotation 
matrix in the $i$--$j$ plane, whose elements 
$[ U_{ij} ] _{pq}$ are defined as
\beq
[ U_{ij}(\theta,\delta)] _{pq} = \left\{
\begin{array}{ll}
\cos \theta & p=q=i \mbox{ or } p=q=j \\
1 & p=q \neq i \mbox{ and } p=q \neq j \\
\sin \theta e^{-i\delta} & p=i \mbox{ and } q=j \\
- \sin \theta e^{i\delta} & p=j \mbox{ and } q=i \\
0 & {\rm otherwise}  \; .\\
\end{array}
\right. 
\label{u-elements}
\eeq
The limit when the sterile neutrino is completely
decoupled -- or when it does not exist -- is obtained
simply by putting the mixing angles $\theta_{14}, \theta_{24}$
and $\theta_{34}$ to zero.
In this limit, ${\cal U}$ in (\ref{u-general}) reduces to
the standard Pontecorvo-Maki-Nakagawa-Sakata (PMNS)
neutrino mixing matrix. Since we are interested only in
oscillation experiments, we neglect any Majorana phases.

We expect $\theta_{14}, \theta_{24}$ and $\theta_{34}$, 
the mixing angles involving the sterile neutrino, to be small. 
Indeed, though the 4$\nu$ analysis of atmospheric 
neutrinos give a rather weak bound of 
$\theta_{24}^2 \approx |{\cal U}_{\mu 4}|^2 < 0.19$
\cite{atm-st-quant}, short baseline disappearance experiments
\cite{dydak} constrain $\theta_{24}^2 < 0.013$,
whereas the short baseline appearance experiments
\cite{bugey,karmen,nomad,miniboone}
give a bound of $\theta_{14}\theta_{24} 
\approx |{\cal U}_{e4} {\cal U}_{\mu 4}| < 0.02$.
The atmospheric neutrino data restrict the deviation 
of $\theta_{23}$ from maximality to be $<0.15$ rad 
\cite{atm-fit},
and the CHOOZ data \cite{chooz} combined with solar,
atmospheric and KamLAND experiments constrain 
$\theta_{13}$ to be less than 0.2 rad \cite{nu04goswami}.
In order to keep track of the smallness of quantities,
we introduce an auxiliary number $\lambda \equiv 0.2$
and define the small parameters to be of the form
$a \lambda^n$. This allows us to perform  
a systematic expansion in powers of $\lambda$.
For the sterile mixing angles, we define
\beq
\theta_{14} \equiv \chi_{14} \lambda \; \quad 
\theta_{24} \equiv \chi_{24} \lambda \; \quad
\theta_{34} \equiv \chi_{34}  \lambda \; \quad ,
\label{sterile-angles-def}
\eeq
whereas for the active mixing angles, we define
\beq 
\theta_{13} \equiv \chi_{13} \lambda \; , \quad
\theta_{23} \equiv \frac{\pi}{4} + \widetilde{\theta}_{23}
\equiv \frac{\pi}{4} + \chi_{23} \lambda \; .
\label{active-angles-def}
\eeq
Here, all the $\chi_{ij}$ are taken to be ${\cal O}(1)$
quantities.
We also treat the solar mixing angle, 
$\theta_{12} \approx 0.6$, as an ${\cal O}(1)$ quantity.
The limits on the other $\theta_{ij}$s mentioned
above translate to
$\chi_{24} < 0.6$, $\chi_{14} \chi_{24} < 0.5$,
$\chi_{23} < 0.75$ and $\chi_{13} < 1$.

In the long baseline neutrino experiments,
the dominating term in flavor conversions oscillates as
$\sin^2[\dma L/(4E)]$.
Owing to the small value of $\dms L/(4E)$, the 
oscillations due to $\dms$ do not have enough time
to develop, and the effect of $\dms$ may be viewed as
a perturbation to the dominating $\dma$ oscillations.
We treat the ratio $\dms/|\dma| \approx 0.03$ as a small
parameter, and define
\beq
\dmsq_{21}/\dmsq_{32} \equiv \zeta \lambda^2 \; .
\label{dmsq-label-def}
\eeq
Note that $\zeta$ is positive (negative) for the normal
(inverted) neutrino mass ordering.

When neutrinos pass through the earth matter, there
are matter effects that give rise to an effective
potential $V_e = \sqrt{2} G_F N_e$ for the electron
neutrino as compared to the other neutrinos by virtue
of the its charged current forward scattering interactions.
Here $G_F$ is the Fermi constant and $N_e$ is the
number density of electrons.
In addition, all the active neutrinos also get an
effective potential $V_n = - G_F N_n / \sqrt{2}$
compared to the sterile neutrino by virtue of 
their neutral current forward scattering reactions.
Here $N_n$ is the number density of neutrons.
For antineutrinos, the signs of $V_e$ and $V_n$ 
are reversed.
The effective Hamiltonian in the flavor basis is then
\beq
H_f \approx  \frac{1}{2E} \left[ {\cal U}_0 
\left( \begin{array}{cccc}
-\dmsq_{21}&0&0 & 0 \\
0&0&0 & 0\\ 
0&0& \quad \dmsq_{32} & 0 \\
0&0&0 &\dmsq_{42} \\
\end{array} \right) {\cal U}_0^\dagger 
+
\left( \begin{array}{cccc}
A_e + A_n & 0 & 0 & 0 \\
0 & A_n&0&0\\
0 & 0& A_n &0\\ 
0 & 0&0&0\\
\end{array} \right)
\right] \; ,
\label{hf-def}
\eeq
where $A_{e (n) } \equiv 2 E V_{e (n)}$,
and ${\cal U}_0$ is the mixing matrix in vacuum,
whose form is given in (\ref{u-general}). 
Let $H_f$ be diagonalized by a unitary matrix ${\cal U}_m$,
such that
\beq
H_D = {\cal U}_m^\dagger \; H_f \; {\cal U}_m \; ,
\label{hf-diag}
\eeq
where $H_D$ is the diagonal matrix.
The elements $[H_D]_{ii}$, being the eigenvalues of $H_f$,
give the relative values of $\tilde{m}_i^2/(2E)$, 
where $\tilde{m}_i$ are the effective masses of the 
interaction eigenstates in matter.
If we assume that the density encountered by the neutrinos
during their passage through the earth is a constant,
the flavor conversion probabilities may be written
in terms of $\tilde{m}_i^2$ and the elements of ${\cal U}_m$ as
\beq
P_{\alpha\beta} \equiv P(\nu_\alpha \to \nu_\beta) =
\left| 
\sum_{i,j} [{\cal U}_m]_{\alpha i} [{\cal U}_m]_{\beta j}^\ast
\exp \left[ i \frac{(\tilde{m}_j^2 - \tilde{m}_i^2) L}{2E} \right] 
\right|^2 \; .
\label{p-ab}
\eeq
This approximation is valid as long as the neutrino trajectories 
do not pass through the core, and the neutrino energy is not 
close to the $\theta_{13}$ resonance energy in the earth.

In order to calculate ${\cal U}_m$, 
it is convenient to work in the basis of neutrino mass
eigenstates in vacuum. The effective Hamiltonian in this
basis is
\barr
H_v & = & {\cal U}_0^\dagger  \;  H_f \; {\cal U}_0 \; 
\nn \\
 & = &  \frac{1}{2E} \left[
\left( \begin{array}{cccc}
-\Delta m^2_{21}&0&0 & 0 \\
0&0&0 & 0\\ 
0&0& \quad \Delta m^2_{32} & 0 \\
0&0&0 & \Delta m^2_{42} \\
\end{array} \right) 
+
{\cal U}_0^\dagger 
\left( \begin{array}{cccc}
A_e + A_n & 0 & 0 & 0 \\
0 & A_n&0&0\\
0 & 0& A_n &0\\ 
0 & 0&0&0\\
\end{array} \right)
{\cal U}_0 \;  \right] \; ,
\label{hv-def}
\earr
which can be diagonalized by the unitary matrix 
$\widetilde{U}$ defined through
\beq
{\cal U}_m = {\cal U}_0 \widetilde{U} \;
\label{utilde-def}
\eeq
such that
\beq
H_D = \widetilde{U}^\dagger \; {\cal U}_0^\dagger \; H_f \; 
{\cal U}_0 \; \widetilde{U} \; =
\widetilde{U}^\dagger \; H_v \; \widetilde{U} \; .
\label{hv-diag}
\eeq
Using the formal representation of 
the elements of ${\cal U}_0$ as well as $\Delta m^2_{21}$
in terms of $\lambda$ 
as shown in eqs.~(\ref{sterile-angles-def}), 
(\ref{active-angles-def}), and (\ref{dmsq-label-def}),  
the matrix $H_v$ can now be expanded formally in powers of
$\lambda$ as
\begin{equation}
H_v= \frac{\Delta m^2_{32}}{2E} 
\left[ h_0+ \lambda h_1 + \lambda^2 h_2
+ {\cal O}(\lambda^3) \right] 
\; .
\label{h-expand}
\end{equation}
The elements of $h_{0,1,2}$ are functions of all the 
neutrino mixing angles, mass squared differences and 
CP violating phases in general; the exact expressions
are given in Appendix~\ref{perturbation}.
All the elements of the matrices 
$h_1$ and $h_2$ are of ${\cal O}(1)$ or smaller,
so that the techniques of time independent perturbation 
theory can be used to calculate the eigenvalues and 
eigenvectors of $H_v$ that are accurate up to 
${\cal O}(\lambda^2)$. 
The complete set of four normalized eigenvectors gives
the unitary matrix $\widetilde{U}$ that diagonalizes
$H_v$ through eq.~(\ref{hv-diag}). Using 
eq.~(\ref{utilde-def}), one obtains 
the unitary matrix ${\cal U}_m$ that diagonalizes $H_f$
through eq.~(\ref{hf-diag}).
The matrix ${\cal U}_m$ and the eigenvalues of $H_v$
(or $H_f$) allow us to calculate the neutrino
flavor conversion probabilities from eq.~(\ref{p-ab}).

The flavor conversion probabilities of neutrinos,
accurate to ${\cal O}(\lambda^2)$, obtained by assuming
the neutrinos to travel through a constant
matter density, are given in Appendix~\ref{perturbation}.
These expressions seem rather complicated.
However, we can make certain approximations that will
simplify these expressions and bring forth some
important physical insights.
Since we are interested in heavy sterile neutrinos,
we may take $|\Delta m^2_{32}| \ll |\Delta m^2_{42}|$. 
Also, since $|\Delta m^2_{32} L/E| \sim {\cal O}(1)$, we have
$|\Delta m^2_{42} L/E| \gg 1$ and the oscillating terms
of the form $\cos(\Delta m^2_{42} L/E)$ may be averaged out.
In the long baseline experiments,
we are interested in the energy range 1--50 GeV.
Even at the higher end of the energy spectrum,
taking the density of the earth mantle to be $\approx 5$
g/cc, we get $A_{e} \approx 2 \times 10^{-2}$ eV$^2$
and $A_{n} \approx - 1 \times 10^{-2}$ eV$^2$
for neutrinos,
so we also approximate $|A_{e,n}| \ll |\dmsq_{42}|$
wherever appropriate.
With these approximations, the neutrino flavor
conversion (or survival) probabilities for an initial
$\nu_\mu$ may be written as
\beqa
P_{\mu e} & \approx &  
2 \theta_{13}^2 \dd^2 \frac{\sin^2 (\Delta_e-\dd)}{(\Delta_e-\dd)^2}
+ {\cal O}(\lambda^3) \; ,
\hspace{8.85cm}
\label{pmue-simp} 
\eeqa
\beqa
P_{\mu\mu}&\approx&\cos^2{\Delta_{32}}
+4 \widetilde{\theta}_{23}^2 \sin^2{\Delta_{32}}
-\Delta_{21} \sin^2\theta_{12} \sin{2 \Delta_{32}} \nn\\
&&+\frac{\theta_{13}^2 \Delta_{32}}{(\Delta_e - \Delta_{32})^2} 
\left\{ -2\Delta_{32}\cos{\Delta_{32}}\sin{\Delta_e}
\sin({\Delta_e-\Delta_{32}})
+ \Delta_e(\Delta_e - \Delta_{32}) \sin{2 \Delta_{32}} \right\}\nn\\
&&-2 \theta_{24}^2\cos^2{\dd}
+2\theta_{24}\theta_{34}\Delta_n\cos{\delta_{24}}\sin{2\dd}
+{\cal O}(\lambda^3) \; ,
\hspace{5.6cm}
\label{pmumu-simp}
\eeqa
\beqa
P_{\mu\tau}& \approx &\sin^2{\Delta_{32}}
-4 \widetilde{\theta}_{23}^2 \sin^2{\Delta_{32}}
+\Delta_{21} \sin^2 \theta_{12} \sin{2 \Delta_{32}} \nn\\
&&+\frac{\theta_{13}^2\Delta_{32}}{(\Delta_e-\Delta_{32})^2}
\left\{ 2\Delta_{32}\sin{\Delta_{32}}\cos{\Delta_e}
\sin{(\Delta_e-\Delta_{32})} 
- \Delta_e(\Delta_e-\Delta_{32})\sin{2 \Delta_{32}}\right\}\nn\\
&&-(\theta_{24}^2+\theta_{34}^2)\sin^2{\dd}
-\theta_{24}\theta_{34} \left( 2 \Delta_n \cos{\delta_{24}}  
+\sin{\delta_{24}} \right)\sin{2 \dd}
+{\cal O}(\lambda^3) \; ,
\hspace{2.5cm}
\label{pmutau-simp}
\eeqa
where we have defined the dimensionless quantities
$\Delta_{ij} \equiv \dmsq_{ij}L/(4E)$ and
$\Delta_{e,n} \equiv A_{e,n}L/(4E)$ for convenience.
The following observations may be made from the above 
expressions:

\begin{itemize}

\item The leading ${\cal O}(1)$ terms are of the form 
$\sin^2 \Delta_{32}$ or $\cos^2 \Delta_{32}$, corresponding 
to the dominating atmospheric neutrino oscillations.
There is no subleading term of ${\cal O}(\lambda)$.

\item For $P_{\mu e}$, there is no sterile contribution
up to ${\cal O}(\lambda^2)$.
Indeed, the leading order sterile contribution to
$P_{\mu e}$ is proportional to $\theta_{24}^2 \theta_{34}^2$,
which is ${\cal O}(\lambda^4)$.

\item In the expression for $P_{\mu\mu}$ or $P_{\mu\tau}$,
the first line contains the leading oscillating term as 
well as the subleading terms due to the deviation of
$\theta_{23}$ from maximality and due to the nonzero
value of $\dmsq_{21}$. 
The next line gives the contribution from 
$\theta_{13}^2$, which matches the one obtained in
\cite{theta13sq}.
The last line contains the contribution from sterile
neutrinos. 
Whereas it is enough to have either 
$\theta_{24}$ or $\theta_{34}$ nonzero for the sterile
mixing to have an effect on $P_{\mu\tau}$, the 
sterile contribution to $P_{\mu\mu}$ will be present
only for nonzero $\theta_{24}$.

\item Only one CP violating phase, $\delta_{24}$, 
is relevant for the flavor conversion probabilities
up to this order. The phases $\delta_{13}$ and $\delta_{14}$
appear only at ${\cal O}(\lambda^3)$ or higher.
In particular, the CP violating terms proportional
to $(\dmsq_{21}/\dmsq_{32}) \theta_{13}$, as given in
\cite{theta13sq}, are absent since they are of
${\cal O}(\lambda^3)$.

\item 
Note that the leading sterile contribution at the long
baseline experiments is found to be at ${\cal O}(\lambda^2)$.
This may be compared with the CP violation in the active
sector, whose leading contribution appears at 
${\cal O}(\lambda^3)$ and the short baseline appearance
experiments, whose positive results would appear 
only at ${\cal O}(\lambda^4)$ or higher. 
The ${\cal O}(\lambda^2)$ sterile contribution to
$P_{\mu\tau}$, which is proportional to $\sin \Delta_{32}$,
is absent in the short baseline
appearance experiments where in general 
$|\Delta_{42}| \sim {\cal O}(1)$ and
$|\Delta_{32}| \ll 1$, so that $\sin\Delta_{32} \approx 0$.

\item When 
$\Delta_e \approx \Delta_{32}$, the $\theta_{13}$
contribution is enhanced due to the factor 
$(\Delta_e - \Delta_{32})^{-2}$. The analytical
approximation is expected to fail in this region since
even the higher order terms in $\theta_{13}$ may become
significant.

\item The analytic expressions are not expected to 
be valid for large $L/E$ where 
$\Delta_{21}$ would become ${\cal O}(\lambda)$ and 
higher order terms in $\Delta_{21}$ would also
contribute to the probability in 
(\ref{pmumu-simp}) and (\ref{pmutau-simp})
at ${\cal O}(\lambda^2)$.

\item The probabilities in (\ref{pmue-simp}), (\ref{pmumu-simp})
and (\ref{pmutau-simp}) do not involve $\Delta m^2_{st}$, and
have no information on whether the mainly sterile neutrino
$\nu_4$ is heavier or lighter than the other three. 
This is due to our approximation of averaging out the
fast oscillations due to $\Delta m^2_{st}$. This
approximation will be more and more accurate as 
$\dmsq_{\rm st}$ increases.

\end{itemize}

The probabilities for the antiparticles are obtained 
simply by replacing $\Delta_{e,n} \to -\Delta_{e,n}$ 
and $\delta_{ij} \to -\delta_{ij}$. 
The sterile contribution to the CP violation is therefore
given by 
\barr
P_{\mu\mu}-P_{\mubar \mubar} & \approx & 
(P_{\mu\mu}-P_{\mubar \mubar})_{3\nu}
+4\theta_{24}\theta_{34}\Delta_n\cos{\delta_{24}}\sin{2\dd} \; ,
\label{pmumu-cp} \\
P_{\mu\tau}-P_{\mubar \taubar} & \approx & 
(P_{\mu\tau}-P_{\mubar \taubar})_{3\nu} 
-4\theta_{24}\theta_{34} \Delta_n \cos{\delta_{24}}\sin{2\dd} 
- 2 \theta_{24}\theta_{34} \sin{\delta_{24}} \sin 2\dd \; . 
\label{pmutau-cp}
\earr
The CP violating contribution of sterile neutrinos to 
$P_{\mu\mu}-P_{\mubar \mubar}$ is entirely from the earth 
matter effects, whereas for
$P_{\mu\tau}-P_{\mubar \taubar}$, the contribution
comes from both the earth matter effects (through the
$\Delta_n$ term) as well as 
the vacuum mixing matrix ${\cal U}_0$ (from the
$\sin \delta_{24}$ term).

For an initial $\nu_e$, the relevant neutrino
flavor conversion probabilities are
\barr
P_{ee}&\approx&
1- 4 \theta_{13}^2 \Delta_{32}^2 
\frac{\sin^2{(\Delta_e - \Delta_{32})}}{(\Delta_e-\Delta_{32})^2}
-2 \theta_{14}^2
+ {\cal O}(\lambda^3) \; ,
\label{pee-simp} \\ 
P_{e\mu} & \approx & 
2 \theta_{13}^2 \Delta_{32}^2  
\frac{\sin^2{(\Delta_e - \Delta_{32})}}{(\Delta_e-\Delta_{32})^2}
+ {\cal O}(\lambda^3) \; ,
\label{pemu-simp} \\
P_{e \tau} & \approx & 
2 \theta_{13}^2 \Delta_{32}^2  
\frac{\sin^2{(\Delta_e - \Delta_{32})}}{(\Delta_e-\Delta_{32})^2}
+ {\cal O}(\lambda^3) \; ,
\label{petau-simp}
\earr
where we have used the approximations 
$|\Delta_{e,n}| \ll |\Delta_{42}|$, and have averaged out terms
that oscillate as fast as $\sin \Delta_{42}$.
The complete expressions accurate to ${\cal O}(\lambda^2)$ 
may be found in Appendix~\ref{perturbation}.
Clearly, sterile neutrinos have no effect at this order
on these probabilities except on $P_{ee}$, 
and there is no sterile contribution to the
CP violation in any of these three channels.

\begin{figure}
\parbox{8cm}{
\epsfig{file=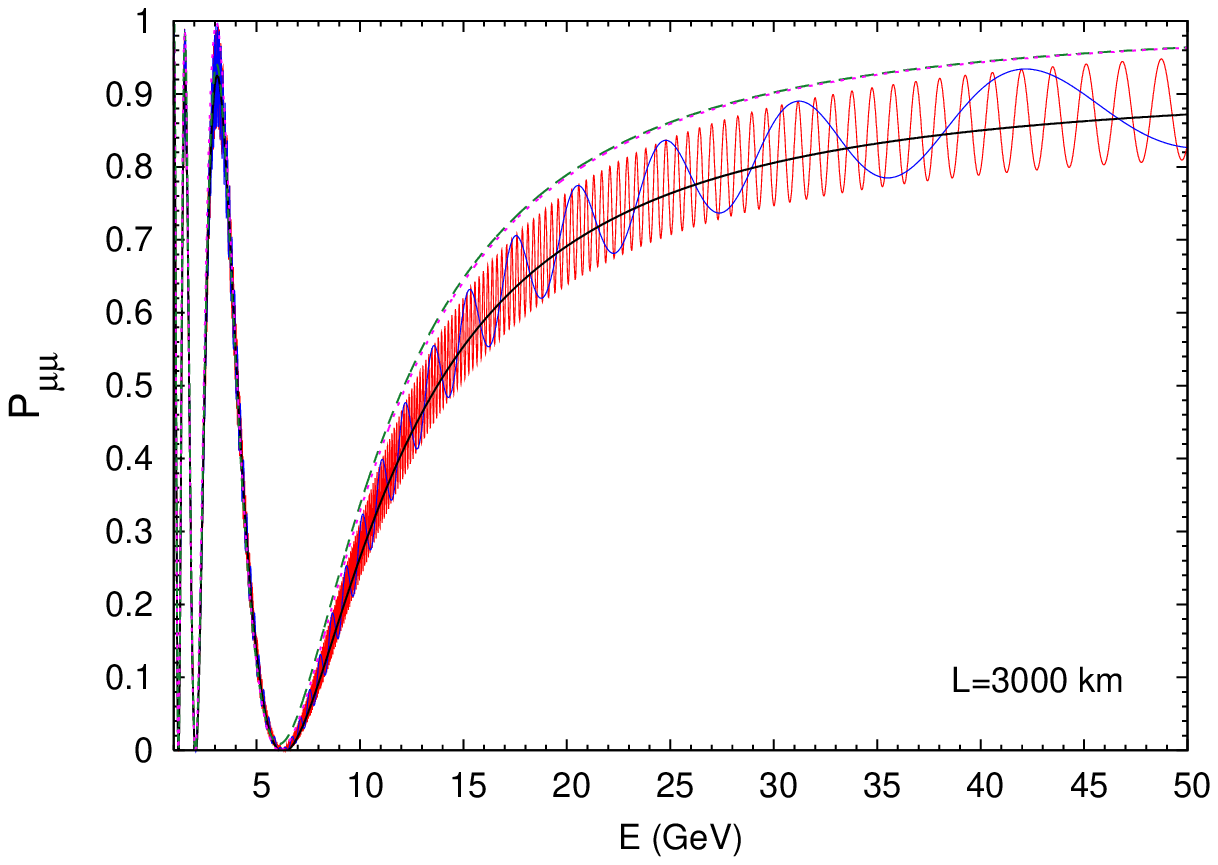,width=7.5cm}
}
\parbox{8cm}{
\epsfig{file=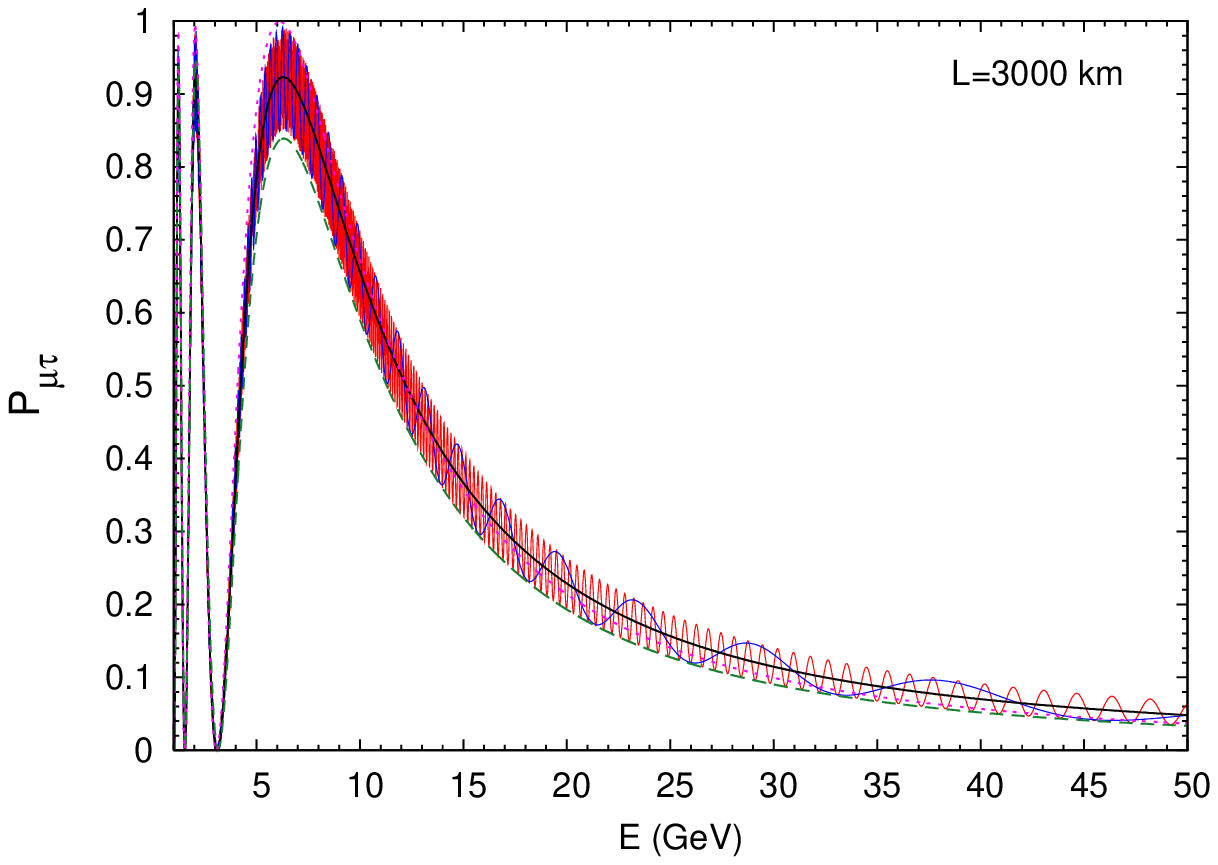,width=7.5cm}
}
\parbox{8cm}{
\epsfig{file=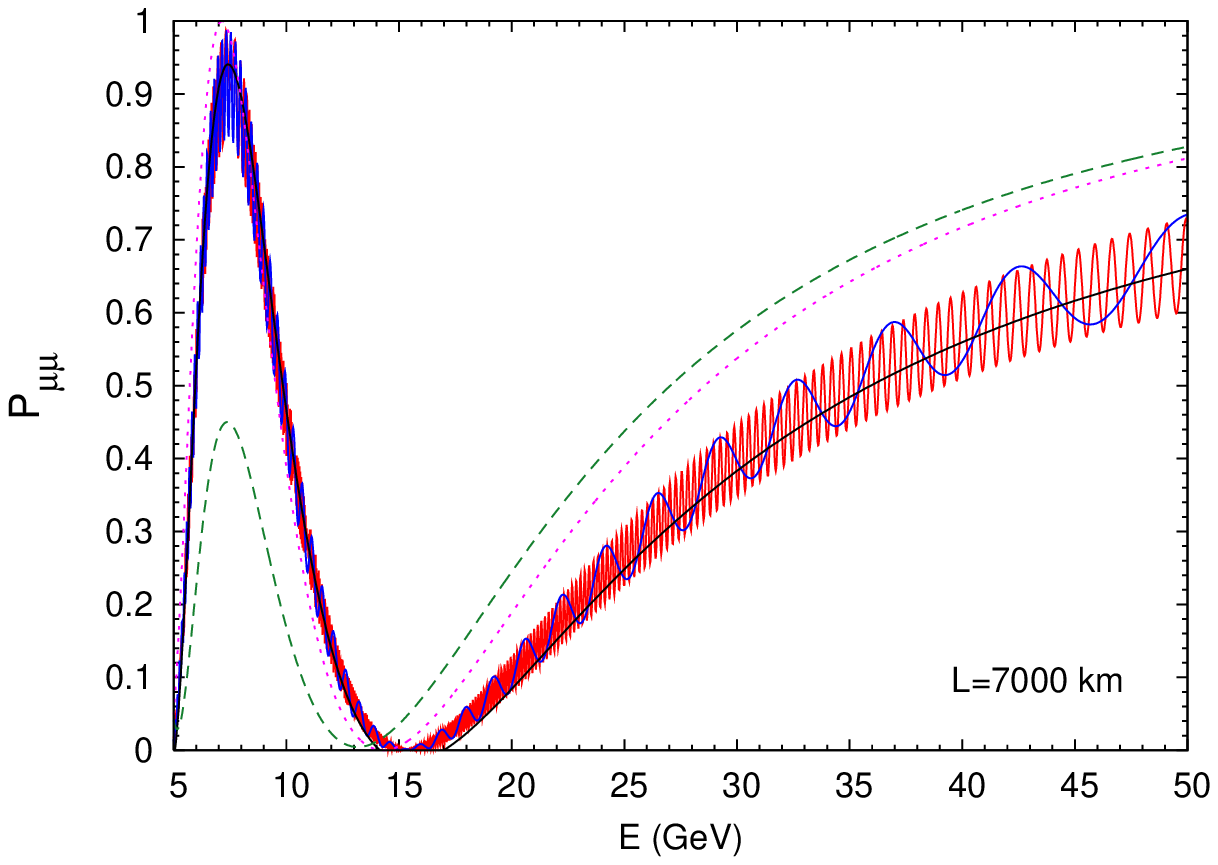,width=7.5cm}
}
\parbox{8cm}{
\epsfig{file=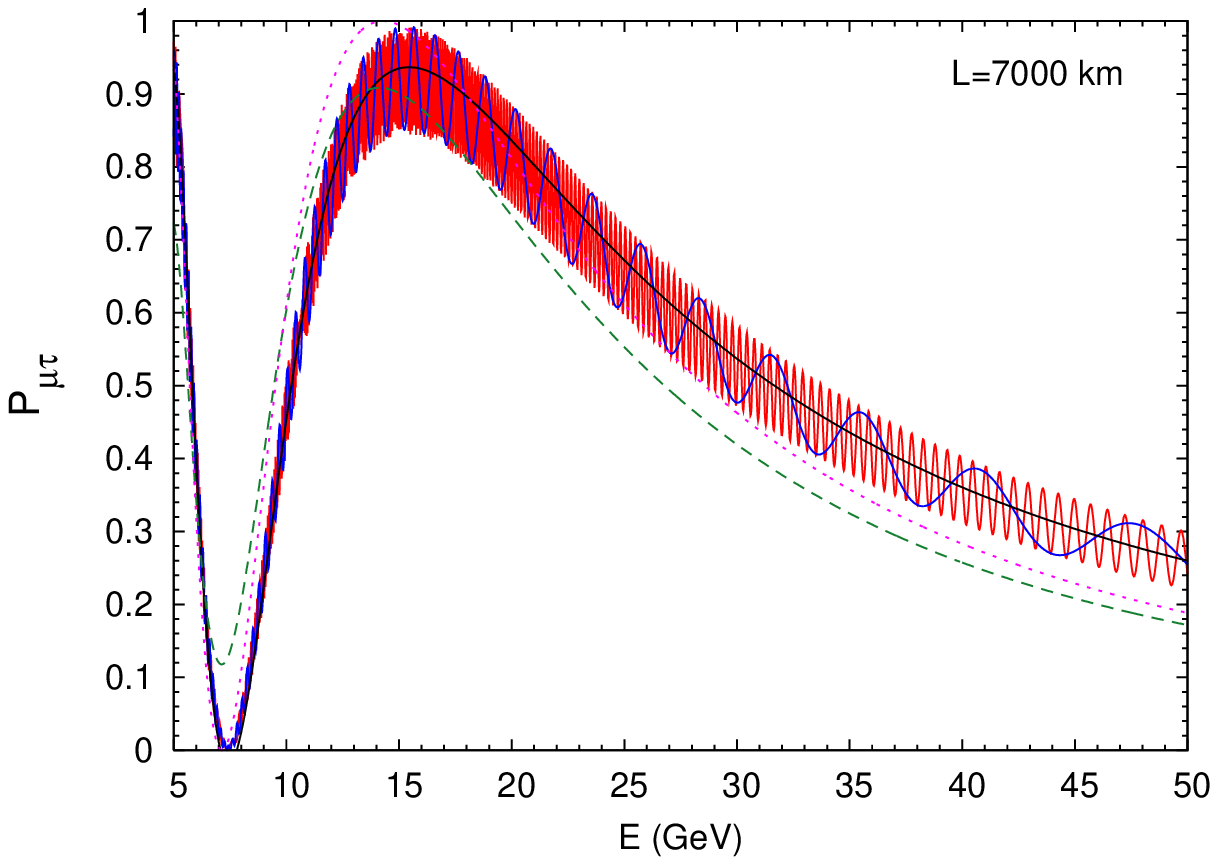,width=7.5cm}
}
\parbox{8cm}{
\epsfig{file=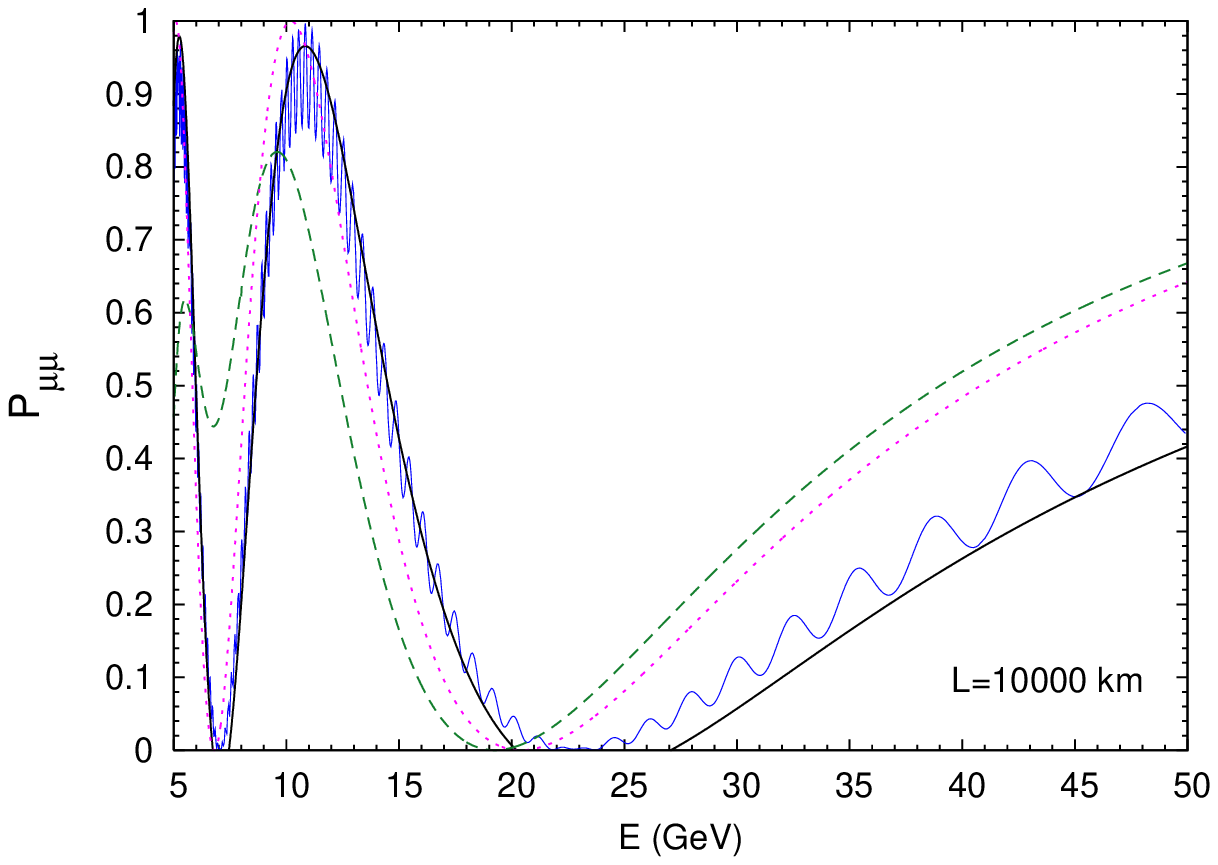,width=7.5cm}
}
\parbox{8cm}{
\epsfig{file=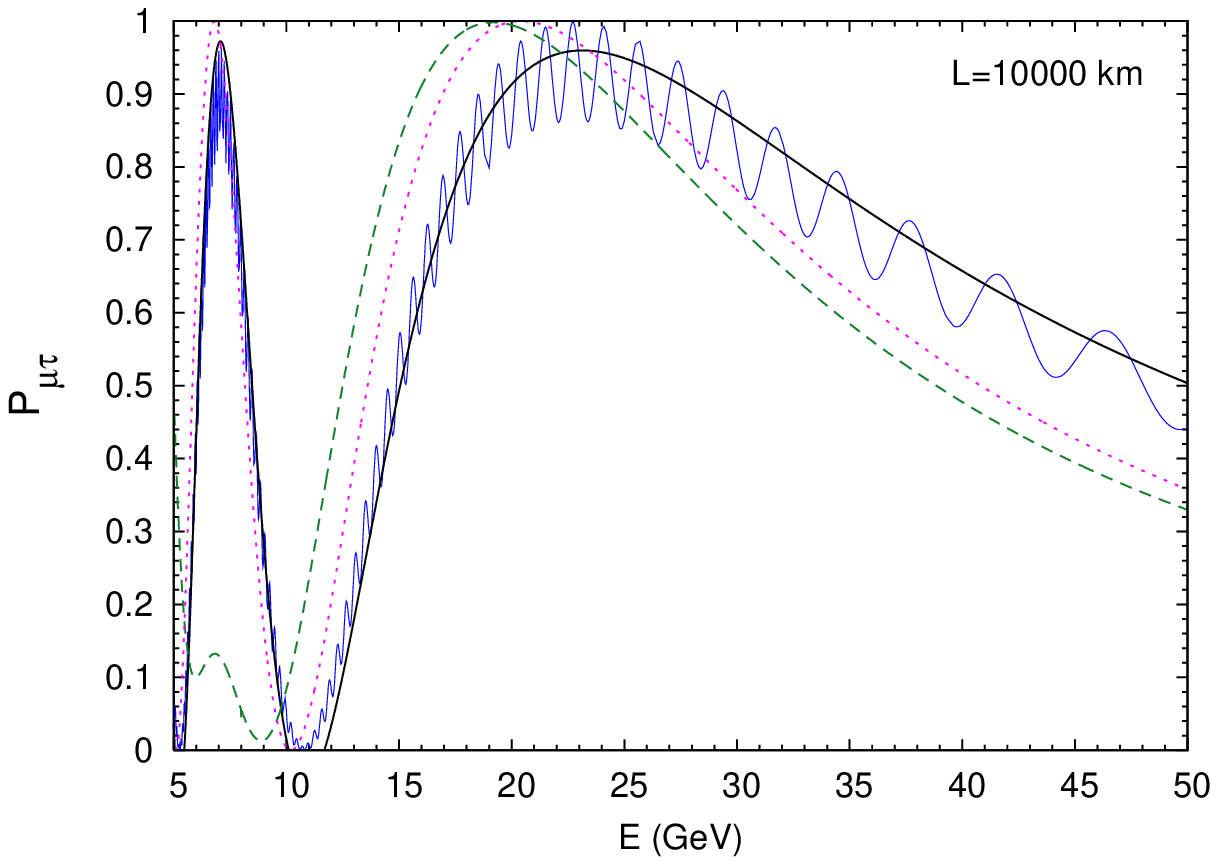,width=7.5cm}
}
\caption{Probabilities $P_{\mu\mu}$ and $P_{\mu\tau}$
as functions of energy, and the comparisons with the
analytic expressions in eqs. (\ref{pmumu-simp}) and 
(\ref{pmutau-simp}). 
In all the plots, we take $\dms = 8 \times
10^{-5}$ eV$^2$, $\dma = 2.5 \times 10^{-3}$ eV$^2$,
$\theta_{23}=45^\circ$, and $\theta_{12}= 33.2^\circ$. 
The magenta (dotted) curve corresponds to the situation with
no sterile contribution and vanishing $\theta_{13}$. 
The blue (red) curve with rapid (extremely rapid) 
oscillations corresponds to $\dmst =$ 0.1 (1.0)
eV$^2$, with $\theta_{14}=\theta_{24} = \theta_{34}=0.2$
rad and $\theta_{13}=0$.
The black curve that passes through the rapidly
oscillating curves denotes the analytical approximation,
which is independent of the value of $\dmst$ since
the high frequency oscillations are averaged out.
The green (dashed) curve represents the situation 
with $\theta_{14}=\theta_{24} = \theta_{34}=0$, but
$\theta_{13}=0.2$ rad.
\label{fig:validity}}
\end{figure}

We demonstrate the validity (and limitations) of our analytic 
approximations in Fig.~\ref{fig:validity}, 
where we show $P_{\mu\mu}$ and $P_{\mu\tau}$ 
as a function of energy for three baselines,
3000 km, 7000 km and 10000 km.
In each panel, we show the probabilities with
$\theta_{14}=\theta_{24}=\theta_{34}=0.2$ rad and
$\theta_{13}=0$, for $\dmst=0.1$ eV$^2$ and 
$\dmst=1$ eV$^2$: the complete 4-neutrino numerical 
simulation with the Preliminary Reference
Earth Model (PREM) \cite{prem}
for the density of the earth, as well as our analytical
approximation that uses the average density along the
path of the neutrino and averages out the high frequency
approximations.\footnote{For a baseline of 10000 km, we only
show $\dmst=0.1$ eV$^2$, otherwise the oscillation
frequency would be too high.}
In order to estimate whether nonzero $\theta_{13}$ can
mimic the signatures of sterile mixing,
we also show the probability for
all the sterile mixing angles vanishing, but 
$\theta_{13}=0.2$ rad.

The following observations may be made:
\begin{itemize}
\item The analytical approximation agrees well with the average
of the exact numerical results for $L = 3000$ km
and 7000 km. 
For $L = 10000$ km, though the analytic approximation 
predicts the qualitative behavior of the averaged 
probabilities, the exact numerical values have an error 
of $\sim 5\%$.
This is due to the large $L$ making $\Delta_{21}
\sim {\cal O}(\lambda)$, so that higher order terms
in $\Delta_{21}^2$ contribute to the probabilities
(\ref{pmumu-simp}) and (\ref{pmutau-simp}).

\item
The dominant effect of the sterile contribution is to 
pull down the value of $P_{\mu\mu}$, which mimics the
deviation of $\theta_{23}$ from its maximal value.
Such a mimicking is also possible through a nonzero
$\theta_{13}$, however the effect of $\theta_{i4}$
may be significantly larger, beyond what is
possible with the current limit on $\theta_{13}$.
Moreover, at energies much larger than the $\theta_{13}$
resonance, the $\theta_{13}$ contribution is suppressed
by the factor $\Delta_{32}/(\Delta_e - \Delta_{32})$ 
in earth matter, whereas the sterile contribution does not
undergo any suppression since $|\Delta_n| \ll |\Delta_{42}|$
in the whole energy range of interest.
One therefore expects that distinguishing the sterile 
contribution would be easier at high energies.

\item
Sterile contribution to $P_{\mu\mu}$ as well as
$P_{\mu\tau}$ is larger at longer baselines, due to the
$\Delta_n$ term present in (\ref{pmumu-simp}) and
(\ref{pmutau-simp}), which increases with increasing
$L$. On the other hand, at low $L/E$ values, the 
sterile contribution to $P_{\mu\tau}$ is highly 
suppressed by the factor 
$\sin \Delta_{32}$ in (\ref{pmutau-simp}).

\end{itemize}

%%%%%%%%%%%%%%%%%%%%%%%%%%%%%%%%%%%%%%%%%%%%%%%%%%%%%%%%
\section{Signatures at long baseline experiments}
\label{lbl}
%%%%%%%%%%%%%%%%%%%%%%%%%%%%%%%%%%%%%%%%%%%%%%%%%%%%%%%%

The analytical expressions 
(\ref{pmue-simp})--(\ref{pmutau-simp}) 
indicate that at $E \gsim 10$ GeV
where $|\Delta_e| \gg |\Delta_{32}|$, the
contribution of the currently unknown $\theta_{13}$
is suppressed by a factor $\sim \Delta_{32}/\Delta_e$.
There is no such suppression for the sterile
contribution, since $|\Delta_{e,n}| \ll |\Delta_{42}|$
for $E < 50$ GeV.
For $E \sim$ 5--10 GeV, the earth matter effects cause
an enhancement of $\theta_{13}$ through the factor
$\Delta_{32}/(\Delta_e - \Delta_{32})$. This energy
range is therefore unsuitable for searching for a
sterile contribution to the conversion probabilities.
At $E<5$ GeV also, since the contribution due to
the currently unknown $\theta_{13}$ is at least of the 
same order as the maximum allowed sterile contribution,
discriminating between $\theta_{13}$ and sterile
contributions to the probabilities would need data from
more than one experiment.
A high energy neutrino experiment is therefore preferred.

In order to demonstrate the capability of future long baseline 
experiments in distinguishing the sterile neutrino
contribution to the neutrino flavor conversion
probabilities, we choose a typical neutrino factory
setup \cite{nufact-setup}, with a 50 GeV muon beam 
directed to a 0.5 kt ``near'' detector 1 km away, and a 
50 kt ``far'' detector 7000 km away.
The detectors may be magnetized iron calorimeters
\cite{ino}, 
which can identify the charge of the lepton produced
from the charged current interaction of the neutrino
or antineutrino.
The number of useful muons in the storage ring
is taken to be $1.066 \cdot 10^{21}$, which corresponds to
approximately two years of running with $\mu^-$ and 
$\mu^+$ each at the neutrino factory, using the 
NuFact-II parameters in \cite{nufactII}. 
We implement the propagation of the neutrinos through
the earth using the 5-density model of the Earth, where the 
density of each layer has been taken to be the average of 
the densities encountered by the neutrinos 
along their path in that layer
with the PREM profile \cite{prem}.  
We take care of the detector characteristics 
using the General Long Baseline Experiment Simulator
(GLoBES) \cite{globes}.
This includes an energy resolution of $\sigma_E/E = 15\%$,
an overall detection efficiency of 75\% for all charged leptons,
as well as additional energy dependent post-efficiencies 
that are taken care of bin-by-bin.
We assume perfect lepton charge identification, and 
neglect any error 
due to wrong sign leptons produced from the oscillations of the
antiparticles. 
These can be taken care of in the complete simulation of the detector
once its detailed characteristics are known.

\begin{figure}
\begin{center}
\parbox{8cm}{
\epsfig{file=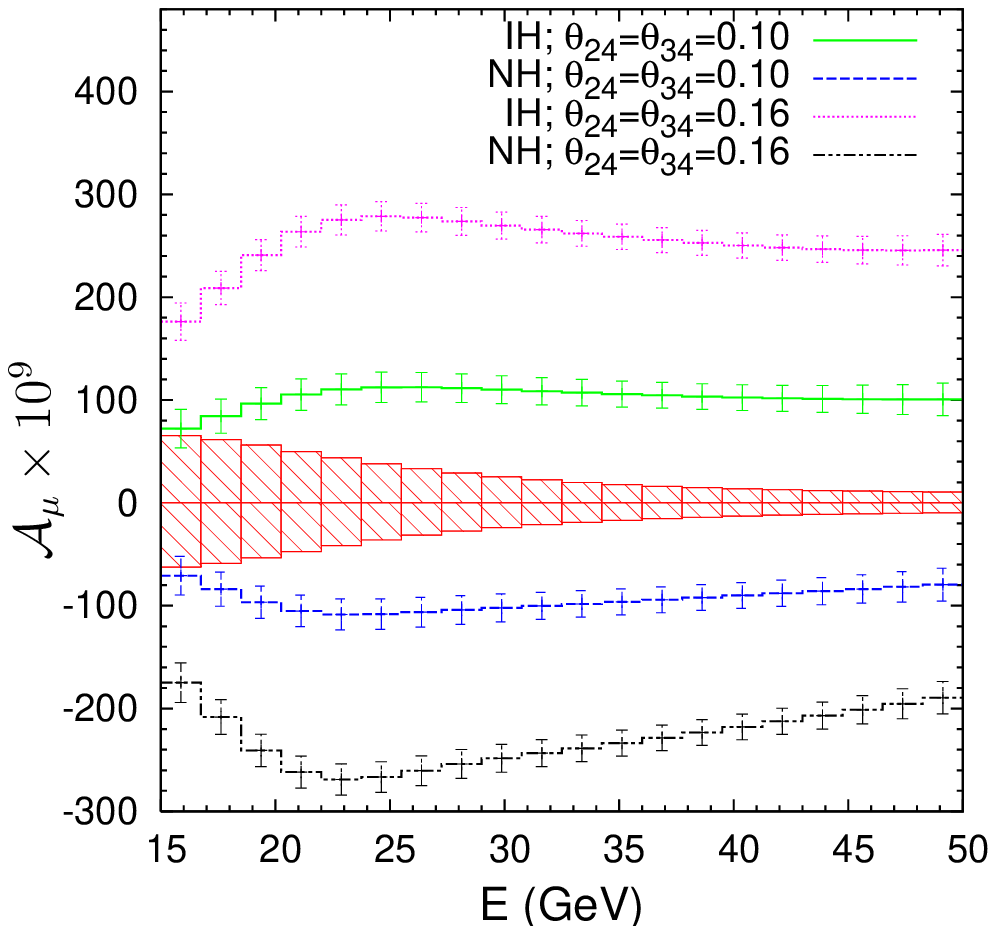,width=7.5cm}
}
\parbox{8cm}{
\epsfig{file=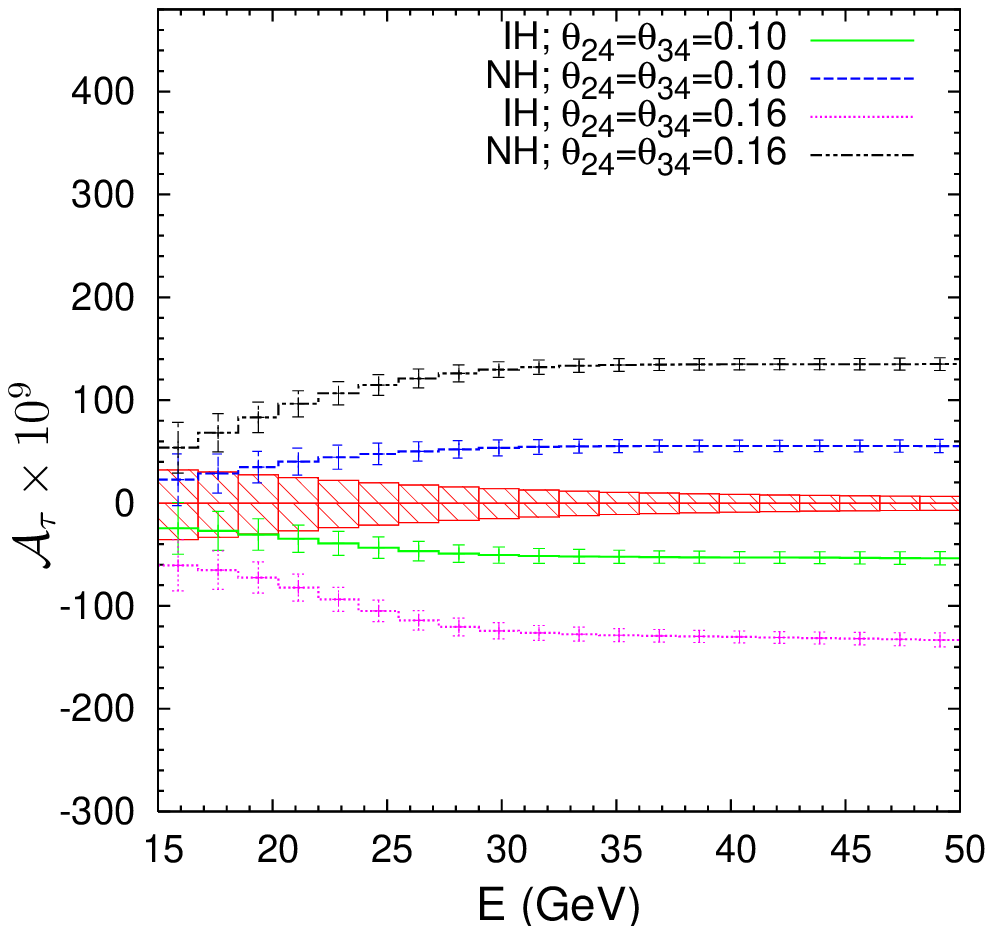,width=7.5cm}
}
\caption {The asymmetries ${\cal A}_\mu(E)$ and
${\cal A}_\tau(E)$ as functions of energy at a
neutrino factory.
The band corresponds to allowed values of the asymmetries
without any sterile mixing, with $\theta_{23}$, $\theta_{13}$ 
and $\delta_{13}$ allowed to vary over all their 
allowed ranges, and with both
the normal (NH) and inverted (IH) hierarchies. 
The plots for showing the dependence on sterile 
components are with $\theta_{23}=\pi/4$, $\delta_{24}=0$, and
$\Delta m^2_{42}=0.1$ eV$^2$. The results will not change 
if $\Delta m^2_{42}$ has higher values. No
significant dependence on $\theta_{14}$ is expected from 
(\ref{pmumu-simp}) and  (\ref{pmutau-simp}),
hence we use $\theta_{14}=0$.
The errors shown are only statistical.
\label{lbl-E}}
\end{center}
\end{figure}

In Fig.~\ref{lbl-E}, we display the asymmetries
\beq
{\cal A}_{\mu}(E) \equiv
\frac{N_\mu^{\rm far}(E)}{N_\mu^{\rm near}(E)} - 
\frac{\nbar_{\mu}^{\rm far}(E)}{\nbar_{\mu}^{\rm near}(E)} \; ,
\quad
{\cal A}_{\tau}(E) \equiv
\frac{N_\tau^{\rm far}(E)}{N_\mu^{\rm near}(E)} - 
\frac{\nbar_{\tau}^{\rm far}(E)}{\nbar_{\mu}^{\rm near}(E)} \; ,
\label{asym-mutau}
\eeq
where $N_\ell$ ($\nbar_{\ell}$) is the number of $\ell^-$
($\ell^+$) observed at the near or far detector. 
These asymmetries roughly correspond to
${\cal A}_\mu \approx P_{\mu\mu} - P_{\mubar \mubar}$ and
${\cal A}_\tau \approx P_{\tau\tau} - P_{\taubar \taubar}$,
where the events observed in the near detector act as a
normalizing factor, and help in canceling out
the systematic errors due to fluxes, cross sections and
efficiencies in each energy bin.
Note that we do not expect any $\tau^\pm$ at the near detector,
hence the number of events of $\tau^\pm$ at the far detector
needs to be normalized to the number of events of $\mu^\pm$ 
at the near detector.

In the absence of any sterile neutrinos, and in the limit of
vanishing $\theta_{13}$, the asymmetries ${\cal A}_\mu$
and ${\cal A}_\tau$ vanish, as can be seen from
(\ref{pmumu-simp}) and (\ref{pmutau-simp}). The $\theta_{13}$
contribution is indeed suppressed at high energies, as 
discussed above.  
In the figure, we show a band corresponding to the possible
signals in the absence of any sterile neutrinos, where
we vary over the allowed values of the angles 
$\theta_{23}, \theta_{13}$, the CP phase $\delta_{13}$
and both the normal as well as inverted mass ordering.
For $\dma, \dms$ and $\theta_{12}$ we only take the 
current best-fit values, since the variation in
these parameters is not expected to cause any significant
change in our results.
We choose to take $\theta_{24}=\theta_{34}$ and $\delta_{24}=0$
for illustration, since from (\ref{pmumu-cp}) and (\ref{pmutau-cp})
we expect the asymmetries to be identical in magnitude and
proportional to the product $\theta_{24} \theta_{34}$ 
with vanishing $\delta_{24}$.
Any discrepancy between these two asymmetries would
indicate a nonzero $\delta_{24}$, and hence CP violation
in the sterile sector.
The third sterile mixing angle, $\theta_{14}$, is taken to be
vanishing since it is not expected to affect the relevant
neutrino conversions.

\begin{figure}
\begin{center}
\parbox{8cm}{
\epsfig{file=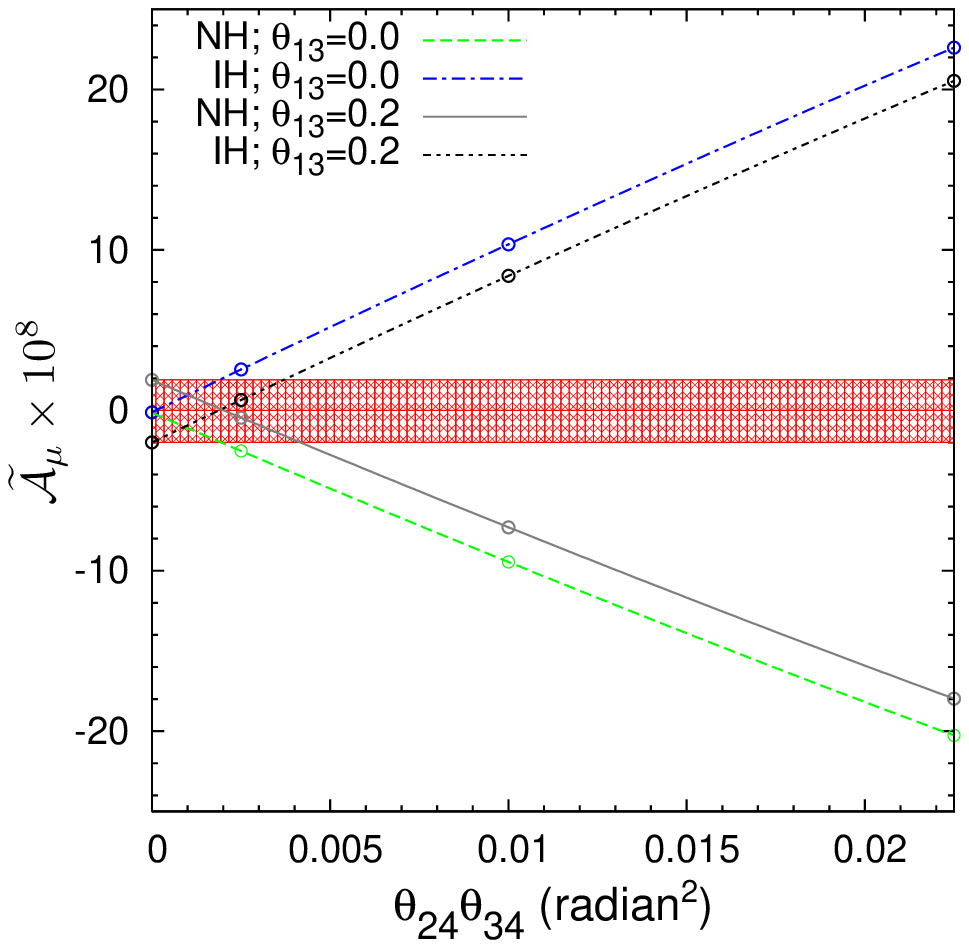,width=7.5cm}
}
\parbox{8cm}{
\epsfig{file=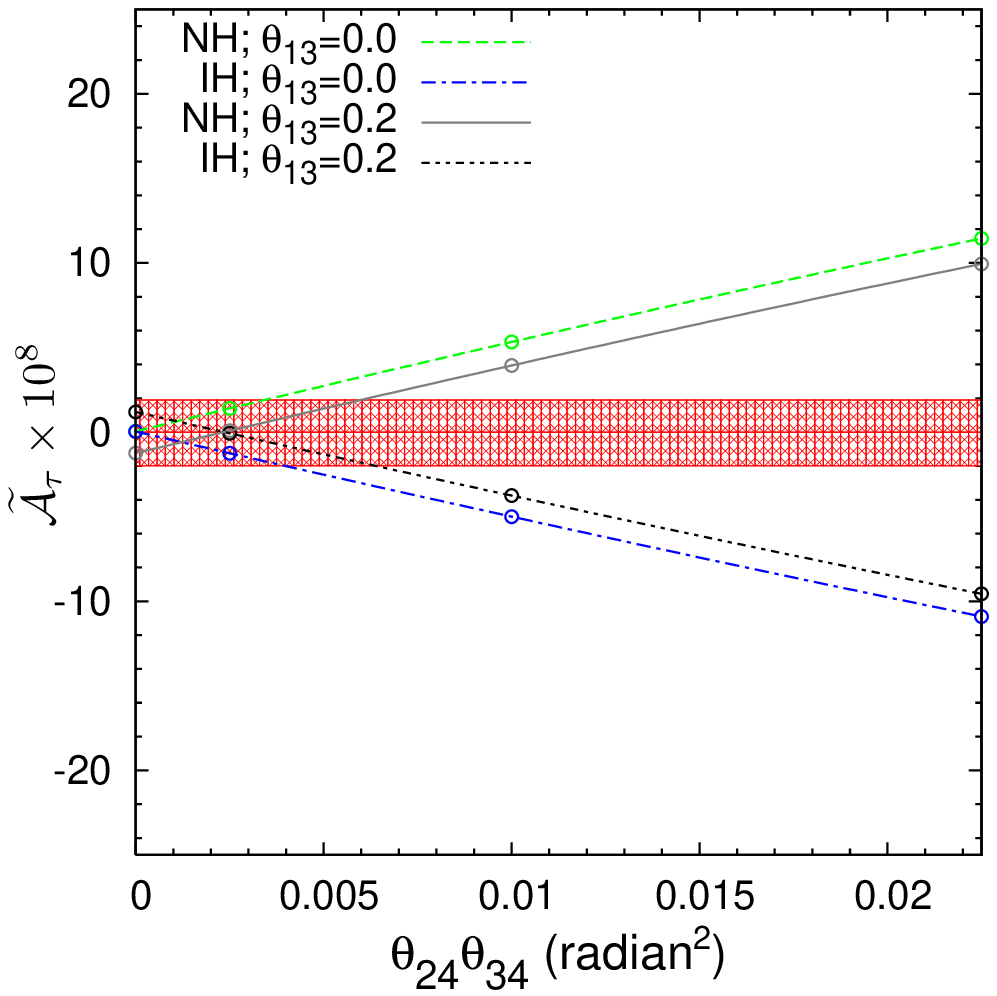,width=7.5cm}
}
\caption {The integrated asymmetries 
$\widetilde{\cal A}_\mu$ and
$\widetilde{\cal A}_\tau$ as functions of sterile
mixing parameters at a neutrino factory.
We use $\theta_{24}=\theta_{34}$ and $\delta_{24}=0$.
The rest of the parameters are the same as in 
Fig.~\ref{lbl-E}.
The statistical errors are smaller than the circles shown
in the plots.
\label{lbl-intE}}
\end{center}
\end{figure}

It may be observed from Fig.~\ref{lbl-E} that for
$E > 15$ GeV, the sterile contribution results in an 
deficit (excess) of the asymmetry for normal (inverted)
hierarchy in the $\mu$ channel. 
In the $\tau$ channel, the situation is the reverse.
This is as expected from our analytic expressions
(\ref{pmumu-cp}) and (\ref{pmutau-cp}).
The asymmetry integrated over energy may
therefore be expected to serve as an efficient discriminator
between the scenarios with and without sterile neutrinos.
In Fig.~\ref{lbl-intE}, we show the integrated asymmetries
\barr
\widetilde{\cal A}_{\mu} & \equiv &
\frac{N_\mu^{\rm far}(E >15 {\rm GeV})}
{N_\mu^{\rm near}(E>15 {\rm GeV})} - 
\frac{\nbar_{\mu}^{\rm far}(E >15 {\rm GeV})}
{\nbar_{\mu}^{\rm near}(E > 15 {\rm GeV})} \; , \nn \\
\widetilde{\cal A}_{\tau} & \equiv & 
\frac{N_\tau^{\rm far}(E> 15 {\rm GeV})}
{N_\mu^{\rm near}(E>15 {\rm GeV})} - 
\frac{\nbar_{\tau}^{\rm far}(E >15 {\rm GeV})}
{\nbar_{\mu}^{\rm near}(E>15 {\rm GeV})} \; .
\label{asym-int-mutau}
\earr
The figure indicates that for $\theta_{24} \theta_{34} \gsim
0.005$, the sterile contribution to neutrino conversions 
can be discernable from the three neutrino mixing results.
The width of the band is determined essentially by the allowed
range of $\theta_{13}$.
If the value of $\theta_{13}$ is bounded further, the 
reach of neutrino factories for the sterile mixing is
enhanced. In addition, the actual value of $\theta_{13}$ also affects
the discovery potential of sterile mixing by influencing
the integrated asymmetries $\widetilde{\cal A}_\mu, 
\widetilde{\cal A}_\tau$, as shown in the figure.
Note that since the asymmetries depend on 
the sign of $\dmsq_{32}$,
sterile mixing also makes it possible to distinguish between 
normal and inverted hierarchies.

\begin{figure}
\begin{center}
\parbox{8cm}{
\epsfig{file=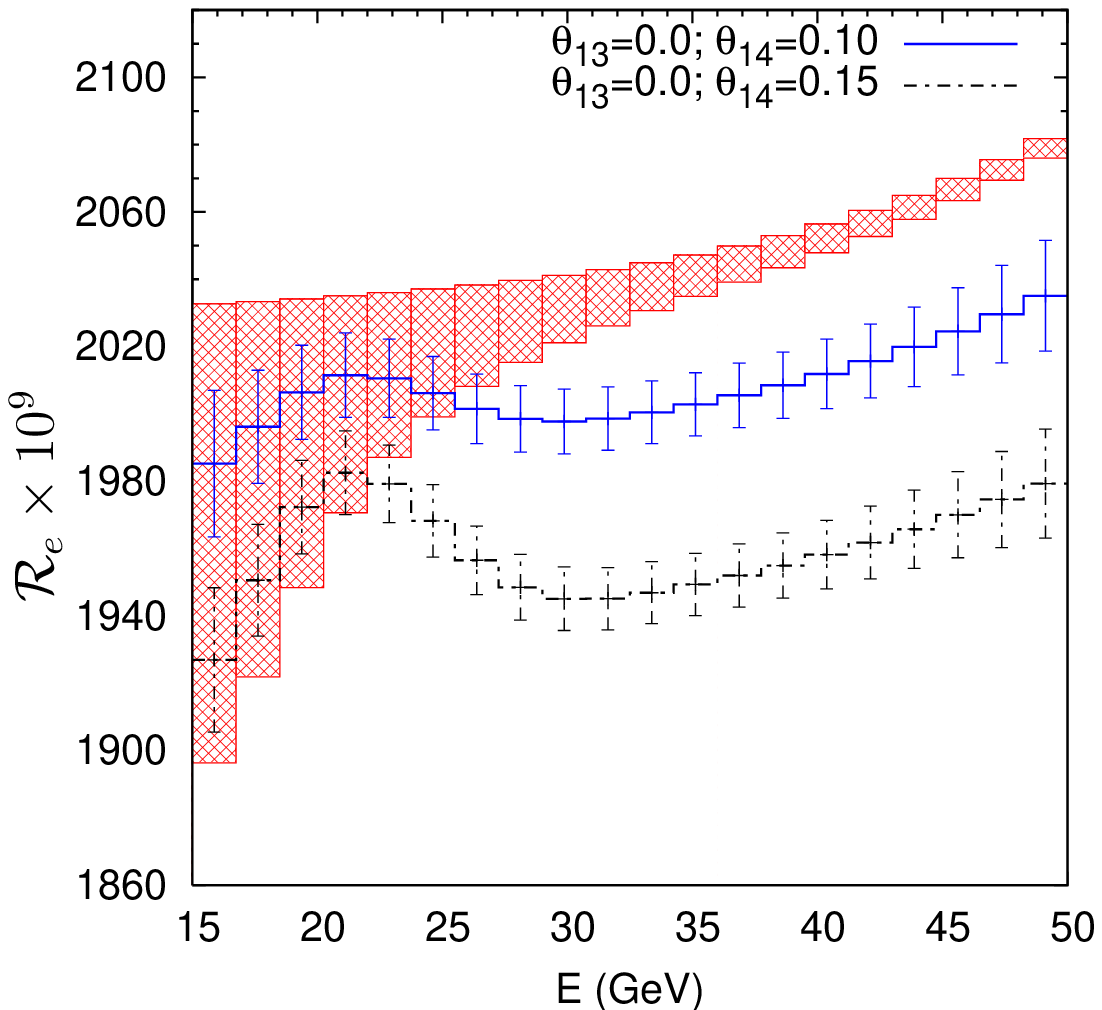,width=7.5cm}
}
\parbox{8cm}{
\epsfig{file=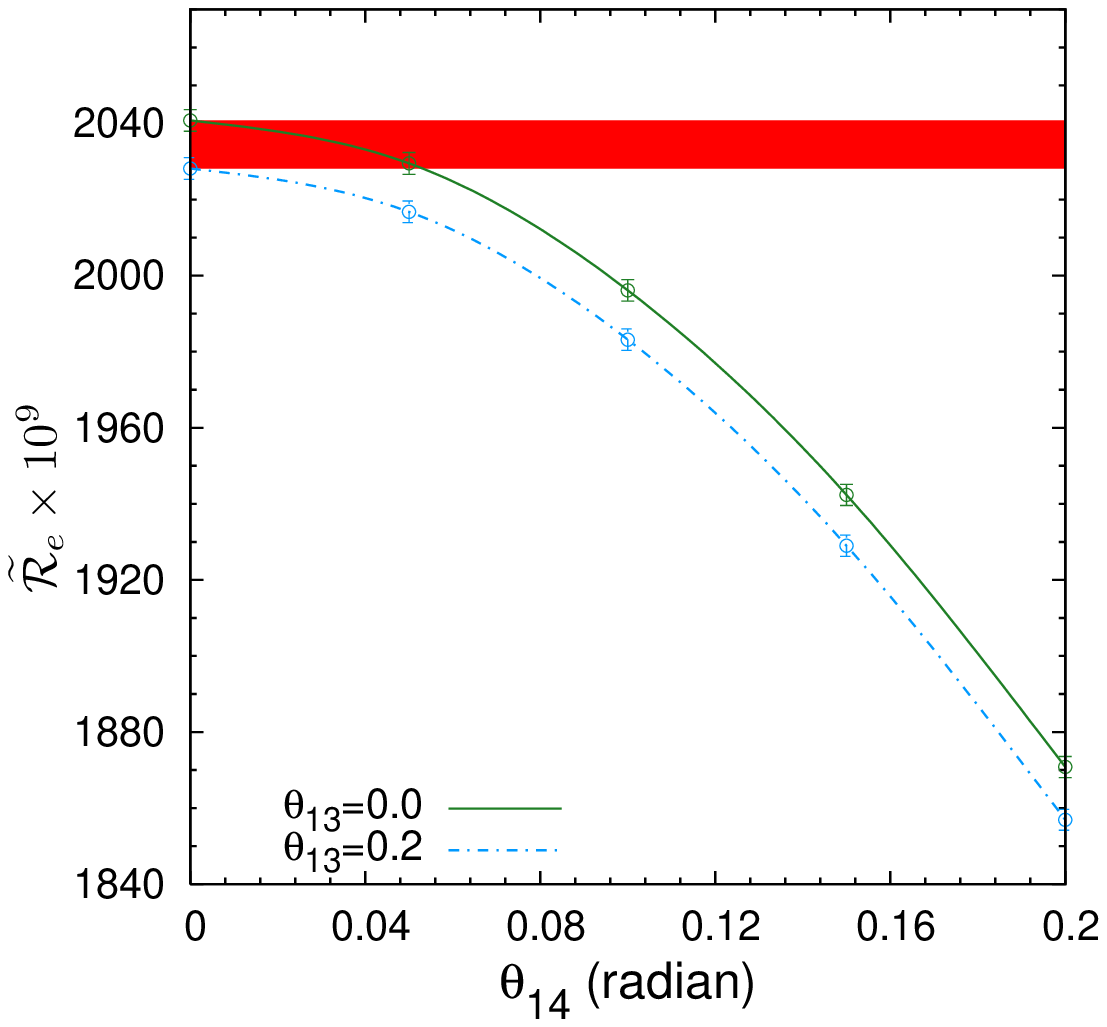,width=7.5cm}
}
\caption{The observables ${\cal R}_e(E)$ and
$\widetilde{\cal R}_e$ at a neutrino factory,
where $e^-/e^+$ and their charge may be identified.
The active neutrino mixing parameters are 
the same as that in Fig.~\ref{lbl-E}. 
The sterile mixing parameters are taken to be 
$\theta_{24}=\theta_{34}=0$ and 
$\Delta m^2_{42}=0.1$ eV$^2$. 
Any increase in $\Delta m^2_{42}$, or
nonzero value of $\theta_{24} /\theta_{34}$ are 
not expected to have any significant effect on this 
observable. The 
result is insensitive to $sgn(\Delta m^2_{32})$. 
\label{lbl-electrons}}
\end{center}
\end{figure}

If we have a 50 kt
detector that can detect $e^-/e^+$ and identify
their charge\footnote{Charge identification 
is needed in order to get rid of the 
error due to misidentification of the wrong sign leptons 
produced due to $\nu_\mu \to \nu_e$ or
$\bar{\nu}_\mu \to \bar{\nu}_e$ oscillations.
A magnetized iron calorimeter with thin iron strips,
or a liquid Ar detector \cite{liq-ar}, may serve the purpose.
If charge identification is not possible,
as in a water Cherenkov detector for example,
the background due to the wrong sign lepton will
have to be taken into account.},
we can use the observable
\beq
{\cal R}_e(E) \equiv 
\frac{N_e^{\rm far}(E)}{N_e^{\rm near}(E)}
\label{pee-E}
\eeq
and the integrated quantity
\beq
\widetilde{\cal R}_e \equiv 
\frac{N_e^{\rm far}(E>25 {\rm GeV})}
{N_e^{\rm near}(E>25 {\rm GeV})}
\label{pee-intE}
\eeq
for detecting the sterile neutrino contribution. Note that
there is no difference between the two hierarchies, or
between $\nu_e$ and $\bar{\nu}_e$, as far as the expected
probabilities are concerned. 
From Fig.~\ref{lbl-electrons}, it may be seen that
for $\theta_{14} \gsim 0.06$, the sterile mixing signals 
can be clearly discerned. 
If the bound on $\theta_{13}$ becomes stronger, even smaller
values of $\theta_{14}$ may be identified.
On the other hand, an higher actual value of $\theta_{13}$
helps in the identification of sterile mixing even at
lower $\theta_{14}$ values.

The ``platinum'' channel $P_{\mu e}$ at the neutrino
factories is not affected by the sterile mixing,
not just to ${\cal O}(\lambda^2)$, but even at 
${\cal O}(\lambda^3)$. Indeed, going to one higher
order in the $\lambda$-perturbation, we get
\begin{eqnarray}
P_{\mu e} &=& 
\frac{\dd^2 \sin^2 ( \Delta_e - \Delta_{32})}{( \Delta_e - \Delta_{32})^2}
\left[
2 \theta_{13}^2 + 
4 \theta_{13}^2 (\widetilde{\theta}_{23} - \theta_{13}) \right] \nn \\
&& + 2 \theta_{13} \Delta_{21} \dd \sin{(2 \theta_{12})}\sin{\dd}
\frac{\cos{(\Delta_e - \dd - \delta_{13})}} {(\Delta_e -\Delta_{32})} \cdot
\frac{\sin \Delta_e}{\Delta_e}
+{\cal O}(\lambda^4) \; .
\end{eqnarray}
For getting $P_{\bar{\mu} \bar{e}}$, one just needs 
to replace $\Delta_e \to -\Delta_e$ and $\delta_{13} \to
-\delta_{13}$.
This channel is therefore not expected to be useful 
in putting constraints on sterile mixing.
On the other hand, it is free of any sterile 
contamination to ${\cal O}(\lambda^3)$,
and is therefore suitable for determining the
parameters in the standard three flavor analysis.

%%%%%%%%%%%%%%%%%%%%%%%%%%%%%%%%%%%%%%%%%%%%%%%%%%%%%%%%%%%%%
\section{Generalization to any number of sterile neutrinos}
\label{many-sterile}
%%%%%%%%%%%%%%%%%%%%%%%%%%%%%%%%%%%%%%%%%%%%%%%%%%%%%%%%%%%%%%

If the LSND results \cite{lsnd} are taken to be valid, 
a single sterile neutrino is not enough to describe all the 
data from  short baseline experiments.
However, two or more sterile neutrinos with
$\dmsq_{j1}\sim 1$ eV ($j>3$) and 
$|{\cal U}_{ej} {\cal U}_{\mu j}| \sim {\cal O}
(0.01$--$0.1)$
are consistent with all data \cite{maltoni-schwetz}.
Some avenues for probing the mixing parameters and
distinguishing between different mass orderings in
such a case
have already been suggested \cite{sn-choubey2,3+2future}.
It is therefore desirable to extend our formalism to 
more sterile neutrinos.

The analytical treatment in Sec.~\ref{expansion} for the
case of one sterile neutrino may be generalized easily to
any arbitrary number $n$ of sterile neutrinos. 
The $(3+n) \times (3+n)$ mixing matrix ${\cal U}$ may be
written in the block form as
\beq
{\cal U} \equiv 
\left( \begin{array}{cc}
{ } [U^{AA}]_{3\times 3} &  [U^{AS}]_{3\times n} \\
{ } [U^{SA}]_{n \times 3} & [U^{SS}]_{n\times n} \\
\end{array} \right)
\equiv
{\cal W} \cdot {\cal V} \equiv
\left( \begin{array}{cc}
{ } [W^{AA}]_{3\times 3} &  [W^{AS}]_{3\times n} \\
{ } [W^{SA}]_{n \times 3} & [W^{SS}]_{n\times n} \\
\end{array} \right)
\left( \begin{array}{cc}
{ } [V^{AA}]_{3\times 3} &  [~0~]_{3\times n} \\
{ } [~0~]_{n \times 3} & [V^{SS}]_{n\times n} \\
\end{array} \right) \; ,
\label{u-nsterile}
\eeq
where $V^{AA} \equiv 
U_{23}(\theta_{23},0) \;
U_{13}(\theta_{13}, \delta_{13}) \;
U_{12}(\theta_{12}, 0)$ is the standard mixing matrix
for three active neutrino flavors, and
$V^{SS}$ is the matrix that mixes the
$n$ sterile neutrinos among themselves.
Since the assignment of ``flavor'' eigenstates to the sterile
species is arbitrary, we choose the basis such that the
flavor and mass eigenstates of the sterile neutrinos
coincide in the absence of any active-sterile mixing,
i.e. $V^{SS} =I_{n\times n}$.
The matrix ${\cal W}$ parametrizes the mixing 
between active and sterile states, and in general may be
represented by a product of matrices $U_{ij}(\theta_{ij},
\delta_{ij})$ as defined in eq.~(\ref{u-elements}), with
$i \leq 3$ and $j>3$.

In addition, we assume that all the active-sterile mixing
is small, which is borne out by the recent 3+2 neutrino fit
to LSND, MiniBOONE as well as the short baseline 
disappearance data \cite{maltoni-schwetz}.
This allows us to write 
\beq
W^{AS} \equiv
\left( \begin{array}{c}
{ } [W^{e S}]_{1\times n} \\
{ } [W^{\mu S}]_{1\times n} \\
{ } [W^{\tau S}]_{1\times n} \\
\end{array} \right)
\equiv \lambda
\left( \begin{array}{c}
{ } [X^{e S}]_{1\times n} \\
{ } [X^{\mu S}]_{1\times n} \\
{ } [X^{\tau S}]_{1\times n} \\
\end{array} \right) 
\equiv
\lambda X^{AS} \; .
\label{x-def}
\eeq
If terms of ${\cal O}(\lambda^3)$ and smaller are neglected, 
the unitary matrix ${\cal W}$ may be written in its most
general form as
\beq
{\cal W} =
\left( \begin{array}{cc}
\left[ I - \lambda^2 \frac{X^{AS} (X^{AS})^\dagger}{2} 
\right]_{3\times 3} &  \left[ \lambda X^{AS} \right]_{3\times n} \\
\left[ - \lambda (X^{AS})^\dagger \right]_{n \times 3} & 
\left[ I - \lambda^2 \frac{ (X^{AS})^\dagger X^{AS}}{2} 
\right]_{n \times n} \\
\end{array} \right) + {\cal O}(\lambda^3) \; .
\label{w-general}
\eeq
The net leptonic mixing matrix ${\cal U}$ in (\ref{u-nsterile})
can then be written as
\beq
{\cal U} = 
\left( \begin{array}{cc}
\left[ \left(I - \frac{W^{AS} (W^{AS})^\dagger}{2} \right)
V^{AA} \right]_{3\times 3} &  
\left[ W^{AS} \right]_{3\times n} \\
\left[ - (W^{AS})^\dagger V^{AA} \right]_{n \times 3} & 
\left[ I - \frac{ (W^{AS})^\dagger W^{AS}}{2} 
\right]_{n \times n} \\
\end{array} \right) + {\cal O}(\lambda^3) \; .
\label{u-net}
\eeq
For the active mixing angles in $V^{AA}$, we use the same
$\lambda$-expansion as in eq.~(\ref{active-angles-def}),
i.e.
$\theta_{13} \equiv \chi_{13} \lambda$ and 
$\theta_{23} \equiv \pi/4 + \widetilde{\theta}_{23}
\equiv \pi/4 + \chi_{23} \lambda$.
We also treat $\dmsq_{21}/\dmsq_{32}$ to be a small
quantity, and denote it formally by
$\dmsq_{21}/\dmsq_{32} \equiv \zeta \lambda^2$, as
in eq.~(\ref{dmsq-label-def}).
The quantities $\zeta, \chi_{ij}$ as well as all the elements of 
$X^{AS}$ are taken to be ${\cal O}(1)$ parameters.

Following the same systematic expansion procedure delineated 
in Sec.~\ref{expansion}  in the case of one sterile neutrino,
we obtain the neutrino flavor conversion (or survival)
probabilities for an initial $\nu_\mu$ beam to be
\beqa
P_{\mu e} & \approx &  
2 \theta_{13}^2 \dd^2 \frac{\sin^2 (\Delta_e-\dd)}{(\Delta_e-\dd)^2}
+ {\cal O}(\lambda^3) \; ,
\hspace{8.65cm}
\label{pmue-simp-n}
\eeqa
\beqa
P_{\mu\mu}&\approx&\cos^2{\Delta_{32}}
+4 \widetilde{\theta}_{23}^2 \sin^2{\Delta_{32}}
-\Delta_{21} \sin^2\theta_{12} \sin{2 \Delta_{32}} \nn\\
&&+\frac{\theta_{13}^2 \Delta_{32}}{(\Delta_e - \Delta_{32})^2} 
\left\{ -2\Delta_{32}\cos{\Delta_{32}}\sin{\Delta_e}
\sin({\Delta_e-\Delta_{32}})
+ \Delta_e(\Delta_e - \Delta_{32}) \sin{2 \Delta_{32}} \right\}\nn\\
&&-2 [W^{\mu S} (W^{\mu S})^\dagger]  \cos^2{\dd}
+2 {\rm Re}[W^{\tau S} (W^{\mu S})^\dagger] \Delta_n \sin{2\dd}
+{\cal O}(\lambda^3) \; ,
\hspace{3.0cm}
\label{pmumu-simp-n}
\eeqa
\beqa
P_{\mu\tau}& \approx &\sin^2{\Delta_{32}}
-4 \widetilde{\theta}_{23}^2 \sin^2{\Delta_{32}}
+\Delta_{21} \sin^2 \theta_{12} \sin{2 \Delta_{32}} \nn\\
&&+\frac{\theta_{13}^2\Delta_{32}}{(\Delta_e-\Delta_{32})^2}
\left\{ 2\Delta_{32}\sin{\Delta_{32}}\cos{\Delta_e}
\sin{(\Delta_e-\Delta_{32})} 
- \Delta_e(\Delta_e-\Delta_{32})\sin{2 \Delta_{32}}\right\}\nn\\
&&- 
\left( [W^{\mu S} (W^{\mu S})^\dagger] 
+ [W^{\tau S} (W^{\tau S})^\dagger] \right)
\sin^2{\dd} \nn \\
& & 
- 2 {\rm Re}[W^{\tau S} (W^{\mu S})^\dagger] \Delta_n \sin{2\dd}
- {\rm Im}[W^{\tau S} (W^{\mu S})^\dagger] \sin{2\dd}
+{\cal O}(\lambda^3) \; .
\hspace{2.75cm}
\label{pmutau-simp-n}
\eeqa
Here we have assumed $|\dmsq_{32}|, |A_{e,n}| \ll |\dmsq_{42}|$,
and have averaged out the oscillating terms of the form
$\cos(\dmsq_{42} L/E)$, as before.
The sterile contribution to the CP violation in these channels
is then
\barr
P_{\mu\mu}-P_{\mubar \mubar} & \approx & 
(P_{\mu\mu}-P_{\mubar \mubar})_{3\nu}
+4 {\rm Re}[W^{\tau S} (W^{\mu S})^\dagger] \Delta_n \sin{2\dd} \; ,
\label{pmumu-cp-n} \\
P_{\mu\tau}-P_{\mubar \taubar} & \approx & 
(P_{\mu\tau}-P_{\mubar \taubar})_{3\nu} 
- 4 {\rm Re}[W^{\tau S} (W^{\mu S})^\dagger] \Delta_n \sin{2\dd}
\nn \\
& & \phantom{(P_{\mu\tau}-P_{\mubar \taubar})_{3\nu} }
- 2 {\rm Im}[W^{\tau S} (W^{\mu S})^\dagger] \sin{2\dd} \; .
\label{pmutau-cp-n}
\earr
For an initial $\nu_e$ beam, the corresponding flavor 
conversion probabilities are
\barr
P_{ee}&\approx&
1- 4 \theta_{13}^2 \Delta_{32}^2 
\frac{\sin^2{(\Delta_e - \Delta_{32})}}
{(\Delta_e-\Delta_{32})^2}
-2 [W^{e S} (W^{e S})^\dagger] 
+ {\cal O}(\lambda^3) \; ,
\label{pee-simp-n} \\ 
P_{e\mu} & \approx & 
2 \theta_{13}^2 \Delta_{32}^2  
\frac{\sin^2{(\Delta_e - \Delta_{32})}}{(\Delta_e-\Delta_{32})^2}
+ {\cal O}(\lambda^3) \; ,
\label{pemu-simp-n} \\
P_{e \tau} & \approx & 
2 \theta_{13}^2 \Delta_{32}^2  
\frac{\sin^2{(\Delta_e - \Delta_{32})}}{(\Delta_e-\Delta_{32})^2}
+ {\cal O}(\lambda^3) \; .
\label{petau-simp-n}
\earr

The mixing matrix ${\cal U}$ in (\ref{u-net}) reduces to
the $4\times 4$ mixing matrix ${\cal U}$ (\ref{u-general}) 
in the case of one sterile neutrino
simply by taking $n=1$ and using the substitution
\beq
W^{e S} \to \theta_{14} e^{-i \delta_{14}} \; , \quad
W^{\mu S} \to \theta_{24} e^{-i \delta_{24}} \; , \quad
W^{\tau S} \to \theta_{34}  \; .
\label{substitution}
\eeq
As a result, the bounds obtained on $\theta_{14}, \theta_{24},
\theta_{34}$ and $\delta_{24}$ in the 4-neutrino analysis 
can be directly translated to bounds on the combinations
$[W^{e S} (W^{e S})^\dagger], [W^{\mu S} (W^{\mu S})^\dagger],
[W^{\tau S} (W^{\tau S})^\dagger]$ as well as the
real and imaginary parts of
$[W^{\tau S} (W^{\mu S})^\dagger]$.
Note that the expressions (\ref{pmue-simp})--(\ref{petau-simp})
obtained in the special case of only one sterile neutrino can
be obtained from the general expressions 
(\ref{pmue-simp-n})--(\ref{petau-simp-n}) 
simply with the substitutions (\ref{substitution}).
Specifically, the bounds obtained on $\theta_{24} \theta_{34}$
in Sec.~\ref{lbl} using the observables 
$\widetilde{\cal A}_\mu, 
\widetilde{\cal A}_\tau$ are simply bounds on
${\rm Re}[W^{\tau S} (W^{\mu S})^\dagger]$.
Similarly, the bound obtained on $\theta_{14}$ through
$\widetilde{R}_e$ is simply the bound on
$[W^{e S} (W^{e S})^\dagger]^{1/2}$.

The above argument also implies that, at least in the region of 
validity of our analytic approximations, the only
combinations of active-sterile mixing parameters
that may be bounded by data are the four quantities
$[W^{e S} (W^{e S})^\dagger], [W^{\mu S} (W^{\mu S})^\dagger],
[W^{\tau S} (W^{\tau S})^\dagger]$ and
$[W^{\tau S} (W^{\mu S})^\dagger]$,
irrespective of the number of sterile species.
For example, in the 3+2 scenario, the mixing matrix ${\cal U}$
may written as
\barr
{\cal U} & = & U_{45}(\theta_{45}, \delta_{45}) \cdot 
U_{35}(\theta_{35}, \delta_{35}) \cdot 
U_{25}(\theta_{25}, \delta_{25}) \cdot  
U_{15}(\theta_{15}, \delta_{15}) \cdot 
U_{34}(\theta_{34}, \delta_{34}) \cdot \nn \\ 
& & \qquad 
U_{24}(\theta_{24}, \delta_{24}) \cdot 
U_{14}(\theta_{14}, \delta_{14}) \cdot 
U_{23}(\theta_{23}, \delta_{23}) \cdot 
U_{13}(\theta_{13}, \delta_{13}) \cdot
U_{12}(\theta_{12}, \delta_{12}) \; ,
\label{u5-allphases}
\earr 
where $\theta_{45}=0$, and
$\theta_{ij} \sim {\cal O}(\lambda)$ for $j>3$.
One may, in addition, choose some of the phases $\delta_{ij}$
to be vanishing by proper redefinitions of leptonic phases.
With the mixing matrix ${\cal U}$ in (\ref{u5-allphases}),
the substitution
\beq
\left( \begin{array}{c}
W^{e S} \\
W^{\mu S} \\
W^{\tau S} \\
\end{array} \right) =
\left( \begin{array}{cc}
\theta_{14} e^{-i \delta_{14}} & \theta_{15} e^{-i \delta_{15}} \\
\theta_{24} e^{-i \delta_{24}} & \theta_{25} e^{-i \delta_{25}} \\
\theta_{34} e^{-i \delta_{34}} & \theta_{35} e^{-i \delta_{35}} \\
\end{array} \right)
\label{substitution-5nu}
\eeq
would give the relevant combinations of the sterile mixing
parameters:
\barr
{ }[W^{e S} (W^{e S})^\dagger] & = & 
\theta_{14}^2 + \theta_{15}^2 \; , \nn \\
{ }[W^{\mu S} (W^{\mu S})^\dagger] & = & 
\theta_{24}^2 + \theta_{25}^2 \; , \nn \\ 
{ }[W^{\tau S} (W^{\tau S})^\dagger] & = & 
\theta_{34}^2 + \theta_{35}^2 \; , \nn \\
{ }[W^{\tau S} (W^{\mu S})^\dagger] & = & 
\theta_{24} \theta_{34} e^{i(\delta_{24}-\delta_{34})} +
\theta_{25} \theta_{35} e^{i(\delta_{25}-\delta_{35})} \; .
\earr
The expected bounds obtained in Sec.~\ref{lbl}
then would correspond to
\beq
\theta_{24} \theta_{34} \cos(\delta_{24}-\delta_{34})
+ \theta_{25} \theta_{35} \cos(\delta_{25}-\delta_{35})
< 0.005 \; , \quad
\sqrt{\theta_{14}^2 + \theta_{15}^2} < 0.06 \; .
\eeq
These bounds will act as a stringent test of the
scenario with multiple sterile neutrinos 
\cite{maltoni-schwetz}.

%%%%%%%%%%%%%%%%%%%%%%%%%%%%%%%%%%%%%%%%%%%%%%%%%%%%%%%
\section{Conclusions}
\label{concl}
%%%%%%%%%%%%%%%%%%%%%%%%%%%%%%%%%%%%%%%%%%%%%%%%%%%%%%%%

Heavy sterile neutrinos may play an important role in
astrophysics and cosmology, for example in r-process
nucleosynthesis or as dark matter. 
Neutrino oscillation experiments, mainly the short
baseline ones, have already put severe constraints 
on the extent of mixing of these sterile neutrinos
with the active ones. 
If the LSND results are taken to be valid, at least
two sterile neutrinos are in fact needed to describe
all data.

Our aim in this paper is to check whether the sterile
neutrinos so constrained can still give rise to 
observable signals at future experiments, and whether
these signals can be cleanly identified in spite of
our current lack of knowledge of all parameters in the
mixing of three active neutrinos.
This would lead to an estimation of bounds on the
sterile mixing parameters that can be obtained with
neutrino oscillation experiments.

The number of neutrino mixing parameters increase
quadratically with the number of neutrinos, and only 
certain combinations are expected to be relevant for 
neutrino flavor conversions.
In order to identify these combinations in an
analytically tractable manner, we exploit the smallness
of certain parameters to carry out a systematic
expansion in an arbitrarily defined small parameter,
$\lambda \equiv 0.2$. The small quantities 
$\theta_{14}, \theta_{24}, \theta_{34}, \theta_{13},
\theta_{23}-\pi/4$, and $\dms/\dma$ are formally
written as powers of $\lambda$ times ${\cal O}(1)$
numbers, and neutrino conversion probabilities 
correct to ${\cal O}(\lambda^2)$ are obtained using
techniques of time independent perturbation theory.
We also neglect terms proportional to $\dma/\dmsq_{42}$,
and average away the fast oscillating terms like 
$\cos(\dmsq_{42} L/E)$ since $|\dmsq_{42} L / E| \gg 1$ 
in typical long baseline experiments.

It is observed that the conversion probabilities 
$P_{\mu e}, P_{e \mu}$ or $P_{e\tau}$ get no sterile
contribution to ${\cal O}(\lambda^2)$. 
For $P_{\mu\mu}$ and $P_{\mu \tau}$, sterile mixing
gives contributions proportional to $\theta_{24}^2$
and $(\theta_{24}^2 + \theta_{34}^2)$ respectively.
In addition, there is a CP violating contribution
proportional to $\theta_{24} \theta_{34}$ to both
these quantities. The survival probability $P_{ee}$
gets modified simply by a term proportional to
$\theta_{14}^2$. There is no dependence on the mass
of the sterile neutrino, since all the terms
containing $\dmsq_{42}$ are averaged out.
It is observed that as long as the neutrinos do not
pass through the core of the earth, the probabilities
obtained through our analytic approximations 
match the exact numerical ones rather well. 
Note that the sterile contribution to the conversion
probabilities at long baseline experiments appears
at ${\cal O}(\lambda^2)$, which is at a lower order than
the appearance of CP violation in the active sector
or the sterile contribution to short baseline
appearance experiments.

Whereas the contribution due to the currently unknown 
$\theta_{13}$ decreases at high energies due to the
earth matter effects, the sterile contribution stays 
almost constant, and therefore the energy range 
$E =$ 10--50 GeV is suitable for distinguishing the 
sterile ``signal'' above the $\theta_{13}$ ``background''.
The CP violating part of the sterile contribution
builds up with increasing $L$, and hence longer baselines
are preferable. This naturally leads to 
the consideration of neutrino
factories with $E_\mu = 50$ GeV and baseline of a 
few thousand km as the desirable setup, with lepton
charge identification capability and a near detector for
calibration purposes.

For illustration we take the far detector to be
near the magic baseline of $\approx$ 7000 km, and
choose three observables, $\widetilde{\cal A}_\mu$ and 
$\widetilde{\cal A}_\tau$ that correspond to the CP asymmetries
in the $\mu$ and $\tau$ channels respectively, and 
$\widetilde{\cal R}_e$, which corresponds to the
disappearance in the electron channel. 
The background in these channels is obtained by
varying over the unknown values of $\theta_{13}, \theta_{23}$
and the CP phase $\delta_{13}$. 
It is observed that the signal rises above this background
for $\widetilde{\cal A}_\mu$ and $\widetilde{\cal A}_\tau$ 
when $\theta_{24} \theta_{34} \gsim 0.005$, and for
$\widetilde{\cal R}_e$ when $\theta_{14} \gsim 0.06$ rad.
The range of $\theta_{i4}$ probed is limited mainly
by the unknown value of $\theta_{13}$.
The limit on $\theta_{13}$ may be brought down by a factor 
of two or more at the reactor experiments like 
Double CHOOZ \cite{dchooz} or Daya Bay \cite{dayabay},
and indeed at the neutrino factories themselves 
\cite{nufact-reach}.
The values of $\theta_{i4}$ that can be probed then
decrease by approximately the same factor.

Note that we have only chosen to analyze a few specific
observables whose dependence on the sterile mixing 
is analytically transparent. A complete analysis that
fits for all the parameters simultaneously may give rise
to more stringent constraints. 
The long baseline experiments thus have the capability of 
tightening the limits on the sterile mixing angles by
almost an order of magnitude over the current ones,
or identify sterile neutrinos if their mixing is indeed
above such a value. Note that if the sterile mixing
is identified through 
$\widetilde{\cal A}_\mu$ or $\widetilde{\cal A}_\tau$,
the neutrino mass hierarchy -- normal vs. inverted --
is also identified.
 
In the light of the recent results that show that LSND,
MiniBOONE and the earlier null-result short baseline
experiments can be consistent if the number of sterile
neutrinos is two or more, we have also extended our formalism 
to include any number of sterile neutrinos. The number
of distinct combinations of sterile mixing parameters
remains the same, irrespective of the number of sterile
neutrinos. We give explicit expressions for such
combinations, and the neutrino conversion probabilities
in terms of them. The limits obtained on $\theta_{i4}$
through the 4-$\nu$ analysis can easily be translated to
the corresponding combinations of these parameters in the
general case. Indeed, the bounds on the sterile mixing parameters
obtained from the measurements described in this paper
would act as stringent tests of the scenarios with
multiple sterile neutrinos.

\section*{Acknowledgements}

We are grateful to P. Huber for his clear
introduction to GLoBES during the JIGSAW07 school and 
W. Winter for clarifying its further details.
We would also like to thank S. Choubey, P. Ghoshal 
and S. Goswami 
for useful discussions and insightful comments.
This work was partly supported through the
Partner Group program between the Max Planck Institute
for Physics and Tata Institute of Fundamental Research.

\appendix

%%%%%%%%%%%%%%%%%%%%%%%%%%%%%%%%%%%%%%%%%%%%%%%%%%%%%%%%%%%%
\section{Flavor conversion probabilities
using perturbation theory to second order}
\label{perturbation}
%%%%%%%%%%%%%%%%%%%%%%%%%%%%%%%%%%%%%%%%%%%%%%%%%%%%%%%%%%%%%

In order to calculate the neutrino conversion (survival)
probabilities in the presence of a sterile neutrino, we
define an auxiliary small parameter $\lambda \equiv 0.2$,
write all the small quantities as $a \lambda^n$ where 
$a$ and $n$ are some constants, and then perform a formal
expansion of the effective Hamiltonian in powers of $\lambda$.
This enables us to use the second order perturbation theory
to get results accurate to ${\cal O}(\lambda^2)$.

We have defined the small quantities in the problem as
\beq
\theta_{14}=\chi_{14} \lambda \; , \quad
\theta_{24}=\chi_{24} \lambda \; , \quad
\theta_{34}=\chi_{34} \lambda \; , \quad
\eeq
\beq
\theta_{13}=\chi_{13} \lambda \; , \quad
\theta_{23}-\pi/4 = \chi_{23} \lambda \; , \quad
\dms/\dma = \zeta \lambda^2 \; .
\eeq

As argued in Sec.~\ref{expansion}, we need to diagonalize
the effective Hamiltonian $H_v$, given in eq.~(\ref{hv-def}).
This Hamiltonian matrix may be expanded in powers of
$\lambda$ as
\beq
H_v = \frac{\dmsq_{32}}{2E} [h_0 + \lambda h_1 + \lambda^2 h_2
+ {\cal O}(\lambda^3)] \; .
\label{h-expand22}
\eeq
Here, the leading term is
\beq
h_0 = \left( \begin{array}{cccc}
a_n + a_e \cos^2 \theta_{12} 
& a_e \cos \theta_{12} \sin \theta_{12} & 0 & 0 \\
 a_e \cos \theta_{12} \sin \theta_{12} 
& a_n + a_e \sin^2 \theta_{12} 
& 0 & 0 \\
0 & 0 & a_n + 1 & 0 \\
0&0& 0 & \quad \sigma \\ 
\end{array} \right) \; ,
\label{h0-def}
\end{equation}
where $a_{e,n} \equiv A_{e,n}/\dmsq_{32}$ and
$\sigma \equiv \dmsq_{42}/\dmsq_{32} \approx \pm \dmst/\dma$. 
We take the neutrinos to be traversing through 
a constant matter density, so that $a_{e,n}$ are
constants.

The subleading term in (\ref{h-expand22}) is
\beq
h_1 = \left( \begin{array}{cccc}
0 & 0
& a_e \chi_{13} \cos\theta_{12}e^{-i \delta_{13}}
& (a_e + a_n) \chi_{14} e^{-i \delta_{14}}\cos \theta_{12} 
- \frac{a_n}{\sqrt{2}} \sin \theta_{12} 
(\chi_{24}  e^{-i \delta_{24}} - \chi_{34}) 
\\
0 & 0 
& a_e \chi_{13} \sin\theta_{12}e^{-i \delta_{13}}
& (a_e + a_n) \chi_{14}  e^{-i\delta_{14}} \sin \theta_{12} 
+ \frac{a_n}{\sqrt{2}} \cos \theta_{12} 
(\chi_{24} e^{-i\delta_{24}}-\chi_{34})
\\
. & . & 0
& \frac{a_n}{\sqrt{2}} (\chi_{24} e^{-i\delta_{24}} + \chi_{34})
\\
. & . & .
& 0 \\   
\end{array} \right) \; .
\label{h1-expand}
\eeq
The matrix $h_1$ is hermitian, so we do not write its lower
triangular elements for the sake of brevity.
Note that all the elements of $h_1$ are ${\cal O}(1)$.

The expression for the matrix $h_2$ in (\ref{h-expand22}) is rather complicated,
we just give its ten independent elements separately here
for the sake of completeness.
The diagonal elements are
\beqa
h_2^{11} &=& - {\Delta_{32}} \zeta -
[ {a_e} \chi_{13}^2 + \left(  {a_e} +  a_n \right) 
\chi_{14}^2 ] {\cos^2  \theta_{12}} -  \frac{a_n}{2} 
 \left( \chi_{24}^2 + \chi_{34}^2 - 2 \chi_{24} \chi_{34} 
\cos \delta_{24} \right)  {\sin^2  \theta_{12}} \nn \\
&& -{\sqrt{2}} \left(  {a_e} +  a_n \right) \chi_{14}[ \chi_{34}
\cos  \delta_{14} - \chi_{24} \cos ( - \delta_{14} +  \delta_{24})]
\sin \theta_{12} \cos \theta_{12} \; , \nn
\eeqa
\beqa
h_2^{22} &=&- [ {a_e}\chi_{13}^2 +\left(  {a_e} +  a_n \right)
\chi_{14}^2 ] {\sin^2  \theta_{12}} - \frac{a_n}{2}(\chi_{24}^2 + \chi_{34}^2 
- 2 \chi_{24} \chi_{34} \cos\delta_{24}) \cos^2 \theta_{12} \nn\\
&& + {\sqrt{2}} \left( {a_e} + a_n \right) \chi_{14} [\chi_{34}
\cos \delta_{14} - \chi_{24} \cos ( -\delta_{14} +  \delta_{24})] 
\sin \theta_{12} \cos \theta_{12} \; , \nn 
\hspace{3.0cm}
\eeqa
\beqa
h_2^{33}&=& - \frac{a_n}{2} \left( \chi_{24}^2 + \chi_{34}^2 
+ 2 \chi_{24} \chi_{34} \cos \delta_{24} \right) + {a_e} { \chi_{13}}^2 \; , 
\hspace{6.8cm} \nn
\eeqa
\beqa
h_2^{44}&=& a_n \left( \chi_{24}^2 + \chi_{34}^2 \right)  + 
\left(  {a_e} +  a_n \right) { \chi_{14}}^2 \; ,
\hspace{9.0cm}
\label{h2-diag}
\eeqa
while the off-diagonal elements are
\beqa
h_2^{12}&=& \left( -[{a_e} \chi_{13}^2 +
\left(  {a_e} +  a_n \right) \chi_{14}^2 ]
+ \frac{a_n}{2}(\chi_{24}^2 + \chi_{34}^2 
- 2 \chi_{24} \chi_{34} \cos\delta_{24}) \right)
\sin \theta_{12} \cos \theta_{12} \nn \\
&& + \frac{\left(  {a_e} +  a_n \right)}{\sqrt{2}} 
\chi_{14} [\chi_{34} \cos \delta_{14} - 
\chi_{24} \cos ( -\delta_{14} +  \delta_{24})] 
\cos 2\theta_{12} \nn \\
& & + i \frac{ \left(  {a_e} +  a_n \right)}{2} 
\chi_{14} [ - \chi_{34} \sin \delta_{14} -
\chi_{24} \sin ( -\delta_{14} +  \delta_{24}) ] \; ,
\hspace{5.5cm}\nn
\eeqa
\beqa
h_2^{13}&=& - \frac{\left( \chi_{24} 
e^{ i \delta_{24}} + \chi_{34} \right)}{2}
[ {\sqrt{2}} \left(  {a_e} +  a_n \right) 
\chi_{14} e^{ -i  \delta_{14}  } \cos \theta_{12} -
a_n \left( \chi_{24} e^{- i \delta_{24}} - \chi_{34}
\right) \sin \theta_{12} ] \; ,
\hspace{0.5cm}\nn
\eeqa
\beqa
h_2^{23}&=& - \frac{\left( \chi_{24} 
e^{ i\delta_{24}} + \chi_{34} \right)}{2}
[ {\sqrt{2}} \left(  {a_e} +  a_n \right) 
\chi_{14} e^{ -i  \delta_{14} } \sin \theta_{12} +
a_n \left( \chi_{24} e^{- i \delta_{24}} - \chi_{34}
\right) \,\cos \theta_{12} ] \; ,
\hspace{0.5cm}\nn
\eeqa
\beqa
h_2^{14}&=& \frac{ a_n} {\sqrt{2}} 
\left( \chi_{24} e^{- i \delta_{24}} + \chi_{34} \right) 
\left( - \chi_{13} e^{-i \delta_{13}} \cos \theta_{12}  + 
\chi_{23} \sin  \theta_{12} \right) \; ,
\hspace{4.6cm}\nn
\eeqa
\beqa
h_2^{24}&=& 
-\frac{ a_n} {\sqrt{2}} 
\left( \chi_{24} e^{- i \delta_{24}} + \chi_{34} \right) 
\left( \chi_{13} e^{-i \delta_{13}} \sin \theta_{12}  + 
      \chi_{23} \cos  \theta_{12} \right) \; ,
 \hspace{4.6cm}\nn
\eeqa
\beqa
h_2^{34}&=& \frac{a_n}{\sqrt{2}}
\chi_{23} \left( \chi_{24} e^{- i \delta_{24}} -\chi_{34} \right)  
+ \left(  {a_e} +  a_n \right) \chi_{13} \chi_{14}     
e^{ i \left( \delta_{13} - \delta_{14} \right)} \; .
\hspace{4.9cm}
\label{h2-offdiag}
\eeqa
Note that all the elements of $h_2$ are ${\cal O}(1)$ or
smaller.
The dependence on $\dmsq_\odot$ appears only at this order,
and only in the element $h_2^{11}$.

Using the above formal expansion of the effective Hamiltonian,
one can compute the eigenvalues and eigenvectors of $H_v$
correct up to ${\cal O}(\lambda^2)$ by using the techniques
of time independent perturbation theory.
The complete set of four normalized eigenvectors gives
the unitary matrix $\widetilde{U}$ that diagonalizes
$H_v$ through eq.~(\ref{hv-diag}).
Using eq.~(\ref{utilde-def}), we can then compute
the unitary matrix $\,{\cal U}_m$ that diagonalizes $H_f$
through eq.~(\ref{hf-diag}).
The matrix $\,{\cal U}_m$ and the eigenvalues of $H_v$
(or $H_f$) allow us to calculate the neutrino
flavor conversion probabilities from eq.~(\ref{p-ab}).
The complete expressions, accurate to 
${\cal O}(\lambda^2)$, are given below.
\beqa
P_{\mu e} &=& 2 \lambda^2 \chi_{13}^2 \dd^2 
\frac{\sin^2 ( \Delta_e - \Delta_{32})}{( \Delta_e - \Delta_{32})^2}
+ {\cal O}(\lambda^3) \; , 
\label{pmue-full} 
\hspace{8.9cm}
\eeqa
\beqa
P_{\mu\mu}&=&\cos^2{\Delta_{32}}+4 \lambda^2 \chi_{23}^2 \sin^2{\Delta_{32}}
-\lambda^2 \zeta  \sin^2{\theta_{12} \Delta_{32} \sin{2 \Delta_{32}}}
\nonumber\\
&&+\frac{\lambda^2 \chi_{13}^2 \Delta_{32}}{(-\Delta_e + 
\Delta_{32})^2} \left\{ -2\Delta_{32}\cos{\Delta_{32}}
\sin{\Delta_e}\sin({\Delta_e-\Delta_{32}})+ 
\Delta_e(\Delta_e - \Delta_{32}) \sin{2 \Delta_{32}} \right\}\nonumber\\
&&+ \lambda^2 \chi_{24}^2 ~ Q_1 
+ \lambda^2 \chi_{34}^2 ~ Q_2
+ \lambda^2 \chi_{24} \chi_{34} \cos{\delta_{24}} ~ Q_3
+ {\cal O}(\lambda^3) \; ,
\label{pmumu-full} 
\hspace{5.55cm}
\eeqa
\beqa
P_{\mu\tau}&=&\sin^2{\Delta_{32}}-4 \lambda^2 \chi_{23}^2 
\sin^2{\Delta_{32}}
+\lambda^2 \zeta \sin^2{\theta_{12} \Delta_{32} \sin{2 \Delta_{32}}}\nn\\
&&+\frac{\lambda^2 \chi_{13}^2\Delta_{32}}{(-\Delta_e+\Delta_{32})^2}
\left\{ -2\Delta_{32}\cos{\Delta_e}\sin{\Delta_{32}}\sin{(\Delta_{32}
-\Delta_e)} + \Delta_e(-\Delta_e+\Delta_{32})\sin{2 \Delta_{32}}\right\}\nn\\
&&+ \lambda^2 (\chi_{24}^2+\chi_{34}^2) ~ Q_4
+ \lambda^2 \chi_{24}\chi_{34}
(\cos \delta_{24} ~ Q_5 + \sin \delta_{24} ~ Q_6)
+ {\cal O}(\lambda^3) \; ,
\hspace{3.8cm}
\label{pmutau-full}
\eeqa
where we have defined
\beqa
Q_1 & \equiv & 
\frac{1}{4(\Delta_n+\Delta_{32}-\Delta_{42})^2 
(-\Delta_n + \Delta_{42})^2}\times \nn \\
&&
\Big\{ - \left( \Delta_n(\Delta_{32}-2 
\Delta_{42})+2\Delta_{42}(-\Delta_{32}+\Delta_{42}) \right)^2 
\cos{2 \Delta_{32}} \nonumber\\
&&
\left. + \left( \Delta_n \Delta_{32}-2(\Delta_n+\Delta_{32})
\Delta_{42} + 2{\Delta_{42}}^2 \right)^2 
\cos{(2 \Delta_n - 2 \Delta_{42})}\right.\nonumber\\
&&
\left.+ 2\Delta_n^2 \Delta_{32}(\Delta_n-
\Delta_{42})(\Delta_n+\Delta_{32}-
\Delta_{42})\sin{2\Delta_{32}}\right.\nonumber\\
&&
 -2 \left( \Delta_n\Delta_{32}-2(\Delta_n+
\Delta_{32})\Delta_{42}+2{\Delta_{42}}^2\right)^2 
\sin^2({\Delta_n+\Delta_{32}-\Delta_{42}})\Big\} \; ,
\hspace{3.6cm}
\label{Q1} 
\eeqa
\beqa
Q_2 & \equiv &
\frac{\Delta_n^2 \Delta_{32}}{2(\Delta_n+\dd-\Delta_{42})^2 
(-\Delta_n + \Delta_{42})^2} \times \nonumber\\
&& 
\Big\{(\Delta_n-\Delta_{42})(\Delta_n+\dd-\Delta_{42})
\sin{2\dd}\nonumber\\
&& 
 -2 \Delta_{32} \cos{\Delta_{32}}\sin{(\Delta_n-
\Delta_{42})}\sin({\Delta_n+\Delta_{32}-\Delta_{42}}) \Big\} \; ,
\hspace{5.9cm}
\label{Q2} 
\eeqa
\beqa
Q_3 & \equiv &
\frac{\Delta_n \left( \Delta_n (\Delta_{32}-2\Delta_{42})
+2\Delta_{42}(-\Delta_{32}+\Delta_{42}) \right)\cos{\Delta_{32}}}
{(\Delta_n+\Delta_{32}-\Delta_{42})^2 (-\Delta_n + 
\Delta_{42})^2}\times\nonumber\\
&& 
\Big\{ 2(\Delta_n-\Delta_{42})(\Delta_n
+\dd-\Delta_{42})\sin{\dd} \nonumber\\ 
&&
+ \dd \left[ -\cos{\dd}+\cos({2\Delta_n+\dd-2\Delta_{42}}) \right] \Big\} \; ,
\hspace{7.2cm}
\label{Q3}
\eeqa
\beqa
Q_4 & \equiv & 
\frac{1}{8(\Delta_n+\dd-\Delta_{42})^2 
(-\Delta_n+\Delta_{42})^2} \Big\{ 4\Delta_n(\dd-
2 \Delta_{42})(\dd-\Delta_{42})\Delta_{42} \nn \\
&& 
 -4(\dd-\Delta_{42})^2\Delta_{42}^2 -
2\Delta_n^2(\dd^2-2\dd\Delta_{42}+2 \Delta_{42}^2)\nn\\
&& 
\left.+2 \Big[ -2\Delta_n(\dd-2\Delta_{42})(\dd-\Delta_{42})
\Delta_{42}+2(\dd-\Delta_{42})^2\Delta_{42}^2\right.\nn\\
&& 
\left.+\Delta_n^2(\dd^2-2\dd
\Delta_{42}+2\Delta_{42}^2) \Big]\cos{2\dd}
\right.\nn\\
&& 
+ \Delta_n\dd \sin \dd \Big[ 
-8\Delta_n(\Delta_n-\Delta_{42})
(\Delta_n+\dd-\Delta_{42})\cos{\dd} 
\nn \\
&& 
-4 \left[ \Delta_n(\dd-2\Delta_{42}) 
+2\Delta_{42}(-\dd+\Delta_{42})\right] 
\sin{(\dd-2\Delta_{42}+2\Delta_n)} \Big] \Big\} \; ,
\hspace{2.8cm}
\label{Q4}
\eeqa
\beqa
Q_5 & \equiv &
\frac{\sin{\dd}}{2(\Delta_n+\dd-\Delta_{42})^2
(-\Delta_n+\Delta_{42})^2} \times \nn \\
& & \Big\{ \Delta_n
[\Delta_n(\dd-2\Delta_{42})+\Delta_{42}(-\dd+\Delta_{42})] 
\times \nn \\
&& 
\qquad \left.\left. [4(\Delta_n-\Delta_{42})
(\Delta_n+\dd-\Delta_{42})
\cos{\dd}+2\dd\sin{\dd}]\right.\right.\nn\\
&& 
\left.\left. + 2 \Big[-2\Delta_n (\dd-2\Delta_{42}) 
(\dd-\Delta_{42}) 
\Delta_{42}+2(\dd-\Delta_{42})^2\Delta_{42}^2 \right.\right.\nn\\
&& 
\qquad  +\Delta_n^2(\dd^2-2\dd\Delta_{42}+2\Delta_{42}^2)
\Big]
\sin{(2\Delta_n+\dd-2\Delta_{42})} \Big\} \; ,
\hspace{4.3cm}
\label{Q5}
\eeqa
\beqa
Q_6 &\equiv & 
\frac{-4 (\dd-\Delta_{42})\Delta_{42}\sin{(\Delta_n-\Delta_{42}) 
\sin\dd}\sin{(\Delta_n+\dd-\Delta_{42})}}
{(\Delta_n+\dd-\Delta_{42})(-\Delta_n+\Delta_{42})} \; .
\hspace{3.3cm}
\label{Q6}
\eeqa
Here we have used the shorthand  
\beq
\Delta_e\equiv \frac{A_e L}{4 E_\nu}\;, \quad 
\Delta_n \equiv \frac{A_n L}{4 E_\nu}\;, \quad 
\Delta_{32} \equiv \frac{\Delta m_{32}^2 L}{4 E_\nu} \;, \quad
\Delta_{42} \equiv \frac{\Delta m_{42}^2 L}{4 E_\nu} \;.
\eeq
The probabilities $P_{e\alpha}$ are
\beqa
P_{ee}&=&
1- 4 \theta_{13}^2 \Delta_{32}^2 
\frac{\sin^2{(\Delta_e - \Delta_{32})}}
{(\Delta_e-\Delta_{32})^2}
- 4 \theta_{14}^2 \Delta_{42}^2  
\frac{\sin^2{(\Delta_e + \Delta_n - \Delta_{42})}}
{(\Delta_e+\Delta_n-\Delta_{42})^2}
+ {\cal O}(\lambda^3) \; ,
\label{pee-full} \\
P_{e\mu}&=&
2 \theta_{13}^2 \Delta_{32}^2 
\frac{\sin^2{(\Delta_e - \Delta_{32})}}
{(\Delta_e-\Delta_{32})^2}
+ {\cal O}(\lambda^3) \; ,
\label{pemu-full} \\
P_{e\tau}&=&
2 \theta_{13}^2 \Delta_{32}^2 
\frac{\sin^2{(\Delta_e - \Delta_{32})}}
{(\Delta_e-\Delta_{32})^2}
+ {\cal O}(\lambda^3) \; .
\label{petau-full}
\eeqa

%%%%%%%%%%%%%%%%%%%%%%%%%%%%%%%%%%%%%%%%%%%%%%%%%%%%%%%%%%%%%%

%%%%%%%%%%%%%%%%%%%%%%%%%%%%%%%%%%%%%%%%%%%%%%%%%%%%%%%%%%%%%%

\end{document}